\documentclass[a4paper, USenglish,cleveref]{lipics-v2021}
\pdfoutput=1

\usepackage{etoolbox}
\newtoggle{tr}
\toggletrue{tr} %

\usepackage{xspace}

\usepackage[utf8]{inputenc}
\usepackage{flushend}

\usepackage{amsmath,amssymb,amsfonts,bm,mathtools}

\usepackage{sidecap}
\sidecaptionvpos{figure}{t}
\usepackage{csquotes}
\usepackage{tabularx}
\usepackage{booktabs}
\usepackage{multirow}
\usepackage{makecell}

\usepackage{pifont}

\usepackage[linesnumbered,noline,noend]{algorithm2e}
\makeatletter
\renewcommand{\@algocf@capt@plain}{above}%
\makeatother

\usepackage{nicefrac}
\usepackage[noend]{algpseudocode}
\algnewcommand{\Label}[1]{\Statex \hspace{-\algorithmicindent}\hspace{-\labelsep} \textbf{#1}}
\algnewcommand{\Input}{\item[\rlap{\textbf{Input:}}{\hphantom{\textbf{Output:}}}]}
\algnewcommand{\Output}{\item[\textbf{Output:}]}
\algnewcommand{\Seq}{\item[\phantom{\textbf{Output:}}]}
\algnewcommand{\IIf}[1]{\State\algorithmicif\ #1\ \algorithmicthen}
\algnewcommand{\EndIIf}{\unskip\ \algorithmicend\ \algorithmicif}
\algnewcommand{\Downto}{\textbf{ downto }}
\algnewcommand{\By}{\textbf{ by }}

\usepackage{csvsimple}

\algnewcommand\algorithmicparfor{\textbf{parfor}}
\algblockdefx{ParFor}{EndParFor}[1]
{\algorithmicparfor\ #1\ \algorithmicdo}
{\algorithmicend\ \algorithmicparfor}

\makeatletter
\ifthenelse{\equal{\ALG@noend}{t}}%
{\algtext*{EndParFor}}
{}%
\makeatother

\usepackage{siunitx}
\makeatletter
\@ifpackagelater{siunitx}{2021/05/17}{%
}{%
	\newcommand{\qty}[2]{\SI{#1}{#2}}
}
\makeatother

\iftoggle{tr}{
	\usepackage[section,above,below]{placeins}
}{
	\usepackage{placeins}
}

\usepackage{hyperref}

\newcommand {\etal}{\mbox{et al.}\xspace} %
\newcommand {\eg}{\mbox{e.\,g.}\xspace}     %

\newcommand{\Oh}[1]{\ensuremath{\mathcal{O}\!\left(#1\right)}\xspace}
\newcommand{\Ohsmall}[1]{\ensuremath{\mathcal{O}(#1)}\xspace}

\newcommand{\Ologn}{\Oh{\log n}}

\newcommand{\Onlogn}{\Ohsmall{n \log n}}
\newcommand{\OnSq}{\Ohsmall{n^2}}

\newcommand{\GitUrl}{https://git.scc.kit.edu/ucsxn/mountains}
\newcommand\tpMap{Tile-Peak map\xspace}
\newcommand{\kirmseAuth}{Kirmse and de~Ferranti\xspace}
\newcommand{\kirmseAlg}{\textsc{ConcIso}\xspace}
\newcommand{\ourAlg}{\textsc{SweepIso}\xspace}
\newcommand{\TNN}{T_{\mathrm{NN}}}
\newcommand{\psfrage}[1]{{\color{blue}{\sf[PS: #1]}}}
\newcommand{\dffrage}[1]{{\color{green}{\sf[DF: #1]}}}
\newcommand{\nhfrage}[1]{{\color{orange}{\sf[NH: #1]}}}
\renewcommand{\psfrage}[1]{}\renewcommand{\dffrage}[1]{}\renewcommand{\nhfrage}[1]{}

\title{A Sweep-plane Algorithm for Calculating the Isolation of Mountains} %

\author{Daniel Funke}{Karlsruhe Institut of Technology}{funke@kit.edu}{}{}%

\author{Nicolai Hüning}{Karlsruhe Institut of Technology}{nicolai.huening@student.kit.edu}{}{}

\author{Peter Sanders}{Karlsruhe Institut of Technology}{sanders@kit.edu}{}{}

\authorrunning{D. Funke, N. Hüning and P. Sanders} %

\Copyright{Daniel Funke, Nicolai Hüning and Peter Sanders} %



\category{} %

\supplementdetails{Software}{\GitUrl} %

\acknowledgements{}%

\nolinenumbers %

\iftoggle{tr}{
\hideLIPIcs %
}

\begin{document}

\maketitle

\begin{abstract}
One established metric  to classify the significance of a mountain peak is its isolation.
It specifies the distance between a peak and the closest point of higher elevation.
Peaks with high isolation dominate their surroundings and provide a nice view from the top.
With the availability of worldwide Digital Elevation Models (DEMs),
the isolation of all mountain peaks can be computed automatically.
Previous algorithms run in worst case time that is quadratic in the input size.
We present a novel sweep-plane algorithm
that runs in time $\Ohsmall{n\log n+p\TNN}$ where $n$ is the input size, $p$ the number of considered peaks and $\TNN$ the time for a 2D nearest-neighbor query in an appropriate geometric search tree.
We refine this to a two-level approach that has high locality and good parallel scalability.
Our implementation reduces the time for calculating the isolation of every peak on earth from hours to minutes while improving precision.
\end{abstract}

\section{Introduction}
High-resolution digital elevation models (DEMs) are an interesting example for large datasets 
with a big potential for applications but equally big challenges due to their enormous size. 
For example, WorldDEM  provided by the TanDEM-X mission covers the entire globe with a resolution of \qty{0.4}{arcsecond}
and is currently the highest-resolution worldwide DEM available \cite{TandemX}.
It consists of $6 \cdot 10^{12}$ individual sample points and amounts to approximately \qty{25}{TB} of data.
Modern LIDAR technology allows $<\qty{1}{\square\meter}$ samples, resulting in more than \qty{300}{TB} of data for the land surface of the earth.
The algorithm engineering community has greatly contributed to unlocking the potential of DEMs by developing scalable algorithms 
for features such as contour lines, watersheds, and flooding risks \cite{contour,watershed,flood}. 
This paper continues this line of research by studying the isolation of mountain peaks 
which is a highly nonlocal feature and requires us to deal with the earths complicated (not-quite spherical) shape.

\begin{figure}[tb]
  \centering
  \includegraphics[width=.33\textwidth]{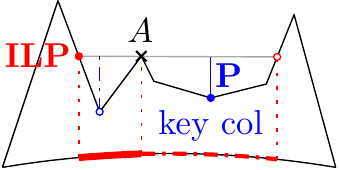}
  \caption{Illustration of a mountain $A$'s isolation (red) and prominence (blue).
  The isolation is measured along the sea-level surface of the earth.}
  \label{fig:isolation}
\end{figure}

A mountain's significance is typically characterized using three properties -- elevation, isolation, and prominence \cite{grimm2004gebirgsgruppen}.
Whereas elevation is a fundamental property, isolation and prominence are derived measures.
Isolation -- also referred to as dominance --
measures the distance along the sea-level surface of the Earth between the peak and the closest point of higher elevation, 
known as the \emph{isolation limit point} (ILP).
Prominence measures the minimum difference in elevation from a peak and the lowest point on a path to reach higher ground -- called the \emph{key col}.
Refer to \cref{fig:isolation} for an illustration.

Whereas previously both measures had to be determined manually by laboriously studying topographic maps,
they can now be computed algorithmically using DEMs.
\kirmseAuth \cite{kirmse2017calculating} present the current state-of-the-art to compute both measures.
They use an algorithm to determine a peak's isolation by searching in concentric rectangles of increasing size around the peak for higher ground.
To determine the isolation of every peak on earth, their algorithm has a worst case running time of \OnSq,
with $n$ being the number of sample points in the DEM.
For high-resolution DEMs of the Earth and other celestial bodies, \eg the Moon \cite{Moon} or Mars \cite{gwinner2010topography},
algorithms with better scalability are needed. 
Also, Continuously Updated Digital Elevation Models (CUDEMs), 
such as NOAA's submarine DEM of North America's coastal regions \cite{NOAA},
require efficient algorithms for frequent reprocessing.
In particular, parallel and external algorithms are required. 

\subparagraph*{Contribution and Outline}
After presenting basic concepts in \cref{sec:prelim} and
related work in \cref{sec:rw} we describe the main algorithmic result in \cref{sec:Algorithm}.
We begin with a sequential algorithm that sweeps the surface of the earth top-down with a sweep-plane
storing the points having surfaces at that elevation.
Processing a peak then amounts to a nearest neighbor query in the sweep-plane data structure.
This results in an algorithm with running time 
$\Ohsmall{n\log n+p\TNN}$ where $n$ is the input size, $p$ the number of considered peaks and $\TNN$ the time for a 2D nearest-neighbor query in an appropriate geometric search tree.
To make this more scalable, we then develop an algorithm working on the natural hierarchy of the data which is specified in \emph{tiles}. This algorithm performs most of its work in two scans of the data which can work independently and in parallal tile-by-tile. Only the highest points in each tile need to be processed in
an intermediate global phase. Each of these three phases has a structure similar to the simple sequential algorithm.

In \cref{sec:SphereTree} we then explain how 2D-geometric search trees can be adapted to work on the surface of the earth by deriving the required 
geometric predicates. 
After outlining implementation details in \cref{sec:impl}, in \cref{sec:Evaluation} we evaluate our approach using the largest publicly available DEM data.

\section{Preliminaries} \label{sec:prelim}

\subparagraph*{Spherical Geometry.}

Planets can generally be approximated by spheres.
In the geographic coordinate system, a point $p = (\phi, \lambda)$ on the surface of a sphere is identified by its latitude $\phi$ and longitude $\lambda$.
Latitude describes $p$'s north-south location, measured from the equator,
$\phi \in (\qty{-90}{\degree}, \qty{90}{\degree}]$ with negative values south of the equator.
Longitude describes $p$'s east-west location, measured from the prime meridian through Greenwich,
$\lambda \in (\qty{-180}{\degree}, \qty{180}{\degree}]$ with negative values west of the prime meridian.

For close distances, Earth's surface is
sufficiently flat to use planar Euclidean
distances between points.  For longer distances,
spherical distance calculations are necessary,
which are computationally more expensive.  On the
surface of the sphere, two points are always
connected by at least two great circle segments.
The shortest such segment, the \emph{geodesic} can
be computed according to
\cref{eq:app:distanceCircle} in the appendix.
As the Earth is not a perfect sphere, it can be
approximated more accurately by an ellipsoid (WGS84 \cite{WGS}), 
which gives rise to the more precise
-- but even more expensive -- distance calculation
according to
\cref{eq:app:distanceEllipsoid}.

\subparagraph*{Digital Elevation Models.}

Digital Elevation Models (DEMs) have become one of the most important tools to analyze the earths surface in geographic information systems.
They represent the surface of the Earth by providing elevation measurements on a grid of sample points.
The data is mostly stored in files of \qty{1}{square\ degree} of coverage each, called tiles.
A tile is addressed by its smallest latitude and longitude.
In order to enable seamless processing of several tiles, each tile stores one sample row/column overlap with its neighbors.
The resolution of DEMs is given by the length of one sample at the equator in arcseconds (\qty{}{\arcsecond}).

State-of-the-art worldwide DEMs, such as WorldDEM, provide a resolution of \qty{.4}{\arcsecond} spacing between sample points
-- about \qty{12}{\meter} at the equator \cite{TandemX}.
Local DEMs, such as the $\text{swissALTI}^\text{3D}$, even have a resolution of only \qty{.5}{\meter} sample point spacing \cite{SwissDEM}.
Freely available DEMs, such as The Shuttle Radar Topography Mission (SRTM) \cite{srtm},
provide a resolution of \qty{3}{\arcsecond} -- about \qty{90}{\meter} at the equator -- 
and an absolute vertical height error of no more than \qty{16}{\meter} for \qty{90}{\percent} of the data \cite{srtmaccuracy}.
Unfortunately, it does not provide global coverage\footnote{%
  Only areas between \qty{60}{\degree} North and \qty{56}{\degree} South are covered.}
and contains large void areas, especially in mountainous regions, which are of particular interest for us.
In a laborious process, de~Ferranti \cite{viewfinderpanoramas} fused raw SRTM data with other publicly available datasets \cite{ASTER,radarsetAntarcticMapping}
and digitized topographic maps to create a worldwide, void-free DEM available at
\url{www.viewfinderpanoramas.org}.
\cref{fig:app:vergleich-srtm} in the appendix illustrates the difference between raw SRTM data and the viewfinderpanoromas DEM.

\section{Related work}\label{sec:rw}

Graff \etal \cite{graff1993automated} are the first to use DEM data to classify terrain into mounts, plains, basins or flats.
As there is no definitive definition of these terrain features, several methods for terrain classification are studied in the literature \cite{torres2019algorithms},
including fuzzy logic \cite{fisher1998mountain,fisher2004helvellyn} or, more recently, deep learning \cite{torres2018deep}.

Prominence has received most of the attention when it comes to algorithmic computation of mountain metrics \cite{helman2005finest,kirmse2017calculating}.
\kirmseAuth \cite{kirmse2017calculating} present the current state-of-the-art regarding isolation and prominence computation.
As their main focus lies on the prominence calculation, they present a rather simple \OnSq time isolation calculation algorithm.

In their algorithm they first calculate potential peaks which are samples that are at least as high as their eight neighboring samples.
Afterwards, a search for the closest higher ground (ILP) is conducted,
where, centered on each peak, concentric rectangles of increasing size are checked to find a sample with higher elevation.
Since the closest higher ground of a peak could also be on a neighboring tile, these might be checked as well.
If a higher ground was found, the distance to this sample is used to constrain the search in neighboring tiles.
If not, tiles in increasing rectangles around the peak-tile are checked until an ILP is found or the complete world has been checked.
Before searching a neighboring tile, the maximum elevation of the tile is checked. When it is smaller than the peak elevation, the tile can be ignored.
Tiles and their maximum elevation are cached, because they need to be loaded rather frequently.
Inside the tile that contains the peak, planar Euclidean distance approximation are used to find an ILP,
for neighboring tiles, distances are computed according to the spherical distance function.
Because a majority of peaks have a small isolation, the ILP is often within the same tile as the peak,
so mostly planar Euclidean distance approximations are used to determine a peak's ILP. 
This reduces computational costs but also the accuracy for peaks with small isolation.
Only in a final step before output is the distance between a peak and its determined ILP calculated using the precise, 
but expensive, ellipsoid distance function.
\kirmseAuth's algorithm is unnamed,
we will refer to it as \kirmseAlg for brevity.

Sweepline algorithms are introduced by Bentley and Ottmann \cite{BenOtt79} to compute line segment intersections.
The technique is generalized by Anagnostou \etal to three-dimensional space \cite{Sweep3D}.
Sweepline algorithms have been applied to various geometric problems in two- and three-dimensional space, 
such as computing Voronoi diagrams \cite{FortuneSL} or route mining \cite{Ships}.

\section{Algorithm} \label{sec:Algorithm}

\begin{figure}[tb]
  \centering
  \includegraphics[width=0.99\textwidth]{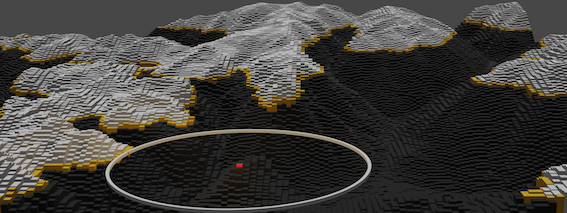}
  \caption{Illustration of DEM grid and sweep-plane algorithm by using a voxel representation.
    Red represents the current peak.
    Orange DEM points are active and contained in the sweep-plane data structure.
    White points are inactive and already removed from the data structure.
    Black points have not yet been processed.}
  \label{fig:illustration-dem-grid-sweepline}
\end{figure}

In this section we present our novel sweep-plane algorithm to calculate the isolation of mountain peaks.
The algorithm takes as input the search area 
-- a quadrilateral $A$ defined by its north-west and south-east corners --
as well as the DEM data.
We will first present a single-sweep algorithm that processes the entire search area in one sweep-plane pass.
Subsequently, we present a scalable three-pass algorithm that
reads the DEM-tiles of $A$ twice and can
process tiles in parallel.

\subsection{Single-Sweep Algorithm}\label{sec:algo:single}

To determine the isolation of a peak, its closest point with higher elevation -- the Isolation Limit Point (ILP) -- needs to be found.
Given the search area $A$,
each sample point $p$ of the DEM within $A$ has two associated events: 
an \emph{insert} event at $p$'s elevation and 
a \emph{remove} event at the elevation of its lowest immediate (NESW) neighbor.
Additionally, if a point is the high-point  of its 3x3 neighborhood it is associated with a \emph{peak} event at its elevation.
We describe the peak detection routine in more detail in \cref{sec:impl}.
All events are created at the beginning of the algorithm and
can therefore be added to a static sequence sorted by descending elevation.
Peak events are processed before other events at the same elevation.

The sweep-plane then moves downward from the highest sample point to the lowest
and traces the contour lines of the terrain,
refer to \cref{fig:illustration-dem-grid-sweepline}.
The sweep-plane data structure $SL$ is a two-dimensional geometric search tree 
that maintains a set of currently \emph{active} points.
A sample point $p$ becomes active at its insert event,
when it is swept by the sweep-plane and inserted into $SL$.
Point $p$ becomes \emph{inactive} and is removed from $SL$ at its remove event, 
at which point its lowest neighboring point is activated
and thus all of $p$'s neighboring points are either active or have already been deactivated.
When a peak event for sample point $p$ is processed,
a nearest neighbor query for the closest active sample point to $p$ in $SL$ is performed.
The returned point $w$ is the ILP for the peak: 
Since $w$ is active and peaks are processed before other events, $w$ must be higher than $p$. 
There cannot be any closer ILP as points activated later are not higher than $p$ 
and since higher points $v$ that are already deactivated are surrounded by points that are all higher than $p$. 
At least one of them must be closer to $p$ than $v$.

\subparagraph*{Analysis.}

Given a DEM with $n$ points, we can detect its $p$ peaks in time $\Oh{n}$.
The resulting $2n + p$ events can be sorted in \Onlogn time by elevation,
Insertion and removal operations take \Ologn worst case time in several geometric search trees \cite{KDTree,QuadTree,smid2000closest}.
Nearest neighbor search complexity is more complicated and still an open problem in many respects.
Therefore, we describe it abstractly as $\TNN$.
Simple trees used in practice such as $k$-D-Trees \cite{KDTree} or Quadtrees \cite{QuadTree} 
achieve logarithmic time ``on average''. Cover trees  \cite{covertree}
achieve logarithmic time when an \emph{expansion parameter} of the input set is bounded by a constant.
Approximate queries within a factor of $1+\epsilon$ are possible in time $\Oh{\log(n)/\epsilon^2}$ \cite{arya1998optimal}.
Overall, we get the claimed time $\Oh{n\log n+p\TNN}$.

\subsection{Scalable Multi-Pass Algorithm}\label{sec:algo:scalable}

\begin{figure}[tb]
	\centering
	\includegraphics[width=.7\textwidth]{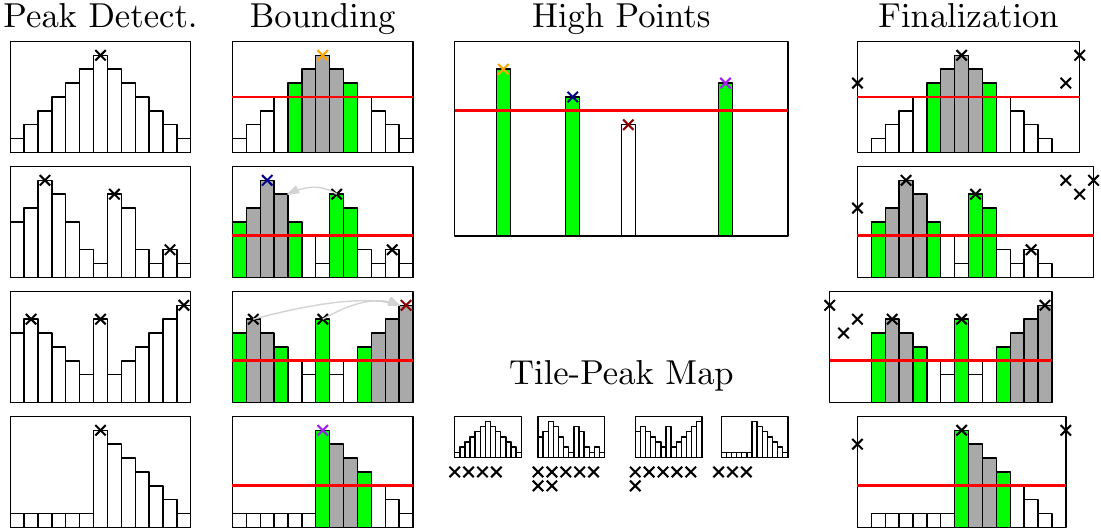}
	\caption{Overview of our scalable multi-pass algorithm for four tiles in two dimensions.
	Peaks are detected for each tile with the algorithm of \kirmseAuth.
	The bounding pass determines an upper bound on the isolation for each peak and assigns it to tiles that could contain closer ILPs in the \tpMap.
	The high-points of each tile without  a local upper bound are processed by a separate pass considering only the high-points.
	In the finalization pass, all peaks assigned to a tile are checked for an ILP within the tile.
	The final determination of the closest ILP out of the found candidates for each peak is not depicted.
	For the sweep-plane passes, green points are active in the sweep-plane data structure, gray points are inactive and hollow points are not yet processed.
	Peaks are marked by crosses.
	The high-point of each tile is marked by a different color cross for each tile. 
}
	\label{fig:alg_overview}
\end{figure}

High-resolution DEMs are massive data sets of currently up to \qty{25}{TB} 
that require scalable algorithms. 
The algorithm described in the previous section can be parallelized to some extend but its sweeping character limits parallelism. Moreover, a geometric search tree covering the entire earth could get quite large.
We therefore develop a two-level algorithm that allows for more coarse-grained parallelism and better locality.
 We adopt the natural hierarchy of the input data using tiles of a fixed area 
 but note that reformatting into smaller or finer tiles would be possible in principle.
Furthermore,  a more general multi-level algorithm could be developed using a similar approach.

Multi-level algorithms start at the finest level
to extract information for global\footnote{Pun intended.} processing at coarser level.
The global results are then passed down
to compute the final solution. In our two-level
algorithm this implies that we have two passes
reading the DEM tiles from external memory while a single global pass
works with simple per-tile informaton. The first (\emph{bounding})
and last (\emph{finalization}) pass can work in parallel on each
tile. The global (\emph{high-point}) pass works in internal memory and is also parallelized.
The first two passes establish a global \tpMap that stores for each tile which peaks \emph{can} have an ILP in it.
The third pass processes these assigned peaks for a tile and determines their ILPs.
All passes follow a very similar structure to our single-sweep algorithm from \cref{sec:algo:single}.
They are described in detail in the following.
Additionally, \cref{fig:alg_overview} provides an overview of our algorithm.

\subparagraph*{Bounding Pass.}
The purpose of the first pass is to establish an upper bound on the isolation of a peak 
and therefore limit the the number of tiles that need to be searched for its ILP in the finalization pass.
To establish this upper bound, we find a tile-local ILP for each peak using our single-sweep algorithm.
Only the highest point in each tile will not have a tile-local ILP and is treated separately in the high-point pass.
Given the upper bound on the isolation of a peak $p$, 
we can assign $p$ to all tiles within radius of the upper bound that could contain a closer ILP for $p$.
In the third pass, these neighboring tiles then need to process $p$ in order to  find ILP candidates within them.
To link peaks to tiles for processing in the third pass, 
we build a \tpMap that stores for each tile a list of peaks that could have an ILP within it.

\subparagraph*{High-Point Pass.}
As there is no local ILP for the high-point $h$ of each tile, we address high-points separately in the second pass.
This pass uses only one type of event -- 
a combination of insert and peak event from the other passes.
We perform a nearest neighbor query in the sweep-plane data structure for $h$'s closest higher point $p$.
The distance between $h$ and $p$ is an upper bound on $h$'s isolation.
The sweep-plane data structure is then traversed again to find all tiles
whose closest point to $h$ is at least as close to $h$ as $p$
and $h$ is linked to these tiles in the \tpMap.
Finally, $h$ is inserted into the sweep-plane data structure and the next lower high-point is processed.

\subparagraph*{Finalization Pass.}
In the third pass, each tile is processed again by a sweep-plane algorithm similar to the single-sweep variant.
The peak events of this pass are all peaks that have been assigned to the tile in the \tpMap.
For each assigned peak, the ILP candidate within the processed tile is determined.
After all tiles have been processed by the third pass,
each peak has as many ILP candidates as tiles it has been linked to.
Thus, in a final step, for each peak the closest found ILP candidate is set as true ILP of the peak.

\subparagraph*{Algorithmic Details and Outline of Analysis.}
Let us first look at the total {\bf work} performed to
process $A$ tiles containing $n$ sample points
overall.  The bounding pass performs a similar
amount of work as the global algorithm except that
it defers nonlocal nearest neighbor searches to
the subsequent passes. More precisely, the high-points of each tile are deferred to the high-point
pass while other peaks are deferred to neighboring
tiles. On average, there is a constant number of such neighbor tiles.%
\footnote{Near the
poles, there are many candidate tiles that may be
closer than a local ILP. However, averaged over
all tiles, this effect is not large. It is
interesting to note though that this is an
artifact of a pseudo-high longitudinal resolution
that is not reflected in the actual precision of
the sensors but stems from artificially mapping
the data using a Mercator projection. Meshes with
more uniform cells are possible, for example the
icosahedral grid used in some modern
climate/wheather models \cite{jungclaus2022icon}.}

The high-point pass potentially defers
nearest-neighbor-search work for the highest
point of each tile to a potentially large number
of tiles. However, this is not so different from
what a search-tree based nearest neighbor search in
a global tree data structure would do. Our two-level algorithm
can be viewed as a vertically split (quad-)tree
algorithm where updates and nearest neighbor
searches are reordered to improve locality. The
overall amount of work done is quite similar
though.

The finalization pass can be viewed as completing
the deferred nearest neighbor searches. Thus, the
main overhead of the two-pass algorithm compared
to the global algorithm is that the data is
sweeped through search trees twice. We mitigate
this effect by building only a coarse search tree
in the bounding pass. This has no negative effect on
precision as the bounding pass is only needed to
identify the tiles where an ILP can be. Only the
finalization pass computes the actual ILPs.

Let us now look at {\bf I/O-costs}. Assume one
tile and the high-points fit in internal memory.
This is similar to a \enquote{semi-external}-assumption used in
previous algorithm engineering papers on DEM processing,
\eg \cite{contour}.
The I/O volume of our two pass algorithm is dominated by
reading all tiles twice.%
\footnote{In comparison, the \tpMap has negiligible data volume and can be handled in an I/O-efficient way by observing that the first two passes only insert to it and the third pass only reads it tile by tile. Thus, we can first buffer its data in an external log which is sorted by tile before the finalization pass.}
This is actually better than an
external-memory implementation of the global
algorithm as that algorithm would have to sort the data by
altitude before being able to scan the input in
the right order for the sweep. Even using
pipelining \cite{DKS08}, sorting would require at
least two reading and one writing pass over the
DEM data.

Finally let us consider {\bf parallelization}.
Bounding and finalization passes can work on $A$
tiles in parallel.
In order to also parallelize 
the high-point pass,
we implement a variant that uses
a static search tree with subtrees augmented by
the maximum elevation occuring in a subtree.  Then
the $A$ nearest-dominating-point searches
can be done in parallel. See
\cref{sec:impl} for details.

\section{Predicates for Search Trees on Spherical Surfaces}\label{sec:SphereTree}

\begin{figure}[tb]
  \centering
  \begin{minipage}{.6\textwidth}
    \begin{subfigure}[t]{.49\textwidth}
      \centering
      \includegraphics[width=\textwidth]{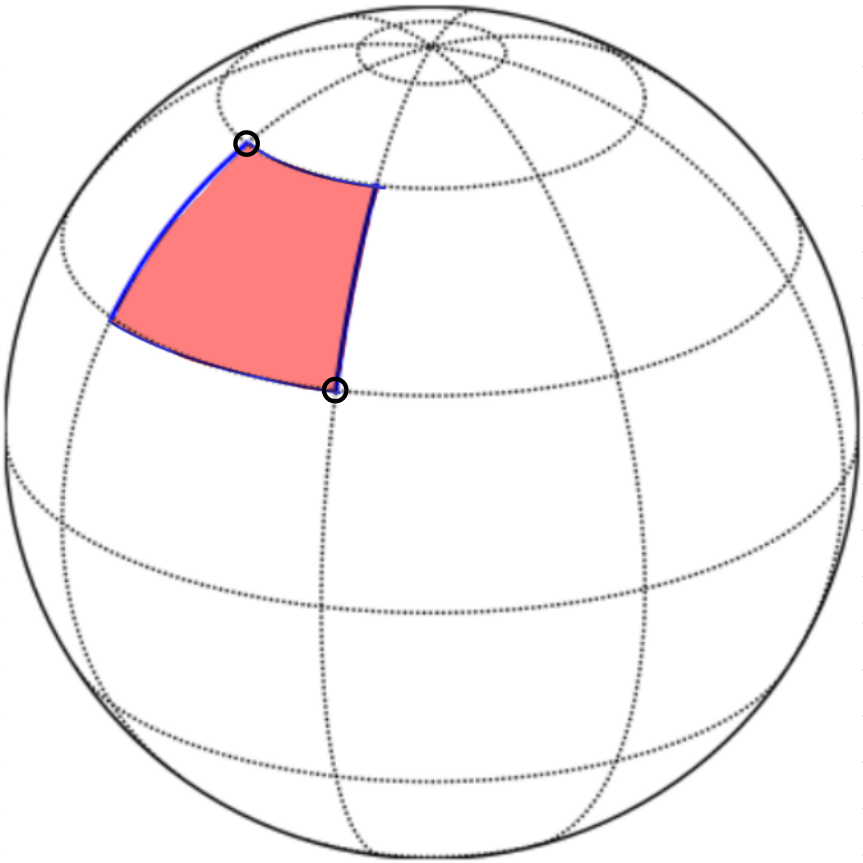}
    \end{subfigure}
    \begin{subfigure}[t]{.49\textwidth}
      \centering
      \includegraphics[width=\textwidth]{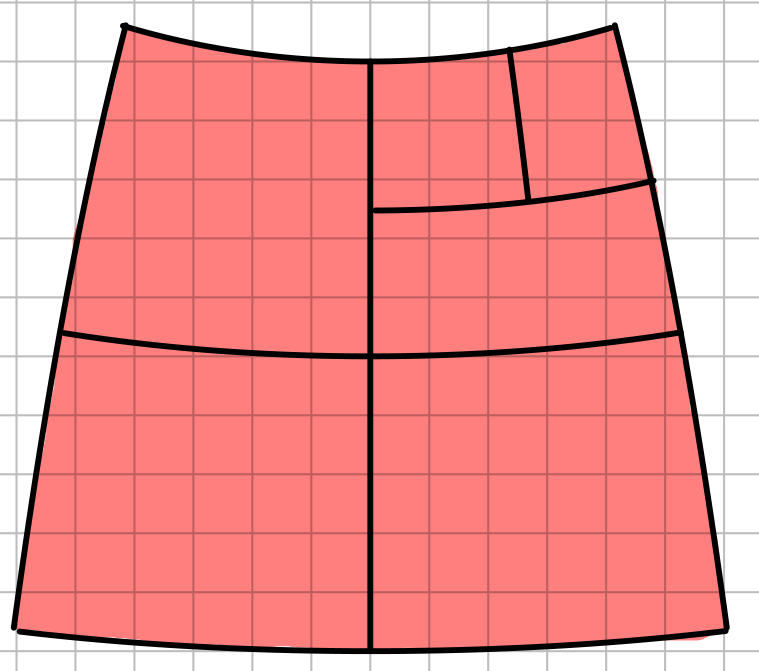}
    \end{subfigure}
    \caption{Representation of a lat-lng aligned quadrilateral.}
    \label{fig:lat-lng-quadrilateral}
  \end{minipage}\hfill
  \begin{minipage}{.3\textwidth}
    \includegraphics[width=\textwidth]{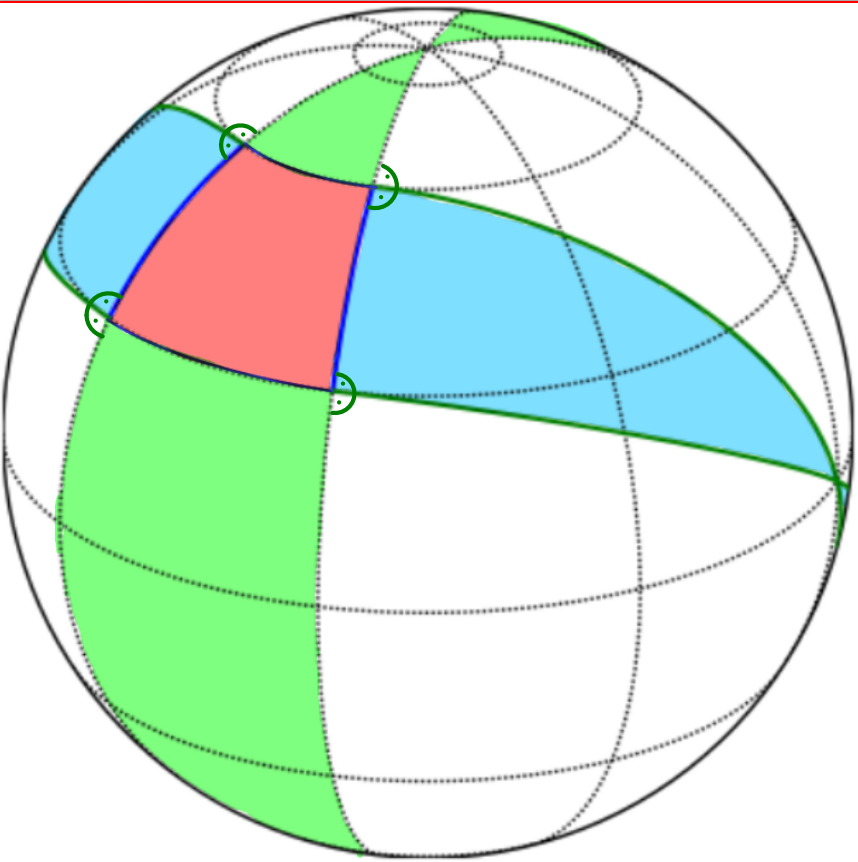}
    \caption{Areas for different min-distance cases between quadrilateral and point.}
    \label{fig:quadrilateral-distance-options}
  \end{minipage}
\end{figure}

The sweep-plane data structure needs to be a dynamic data structures which support efficient nearest neighbor queries.
Space-partitioning trees such as $k$-D trees or Quadtrees are well-suited data structures for this application \cite{KDTree,QuadTree}.
These trees recursively divide the input space into smaller and smaller blocks.
For a spherical surface, the space is divided into quadrilaterals which are aligned with latitude and longitude, refer to \cref{fig:lat-lng-quadrilateral}.
A quadrilateral is defined by its north-west and south-east corners.
Each quadrilateral can then be further subdivided into smaller quadrilaterals.
The root of a space-partitioning tree covers the entire input area.

Our sweep-plane algorithm requires two geometric primitives for these trees:
a) whether given a point $p$ lies inside a quadrilateral $Q$ and b) the shortest distance between $p$ and $Q$.
The former can be easily answered by comparing $p$'s latitude and longitude with $Q$'s north-east and south-west corner.
For the latter,
there are four configurations of $p$ and $Q$ that need to be considered, refer to \cref{fig:quadrilateral-distance-options}:
\begin{enumerate}
  \item $p \in Q$ (red area), \label{distancequadr:case1}
  \item $p$ is between the longitude lines of $Q$ (green area), \label{distancequadr:case2}
  \item $p$ is between the four great circles through the corners of $Q$, which are perpendicular to the longitude edges of $Q$ (blue area), \label{distancequadr:case3}
  \item all other positions (white area). \label{distancequadr:case4}
\end{enumerate}
For case \ref{distancequadr:case1} the distance is zero.
In case \ref{distancequadr:case2}, the closest point $s \in Q$ to $p$ 
is on the intersection between the longitude line of $p$ and one of the latitude-edges of the quadrilateral,
as the shortest distance between two latitudes is along the longitude lines
-- also refer to \cref{fig:app:distance-quadrilateral} in the appendix.
The shortest distance of $p$ to $Q$ is thus the shorter distance between $p$ and the north and south latitude of $Q$.

In the remaining cases, $s$ must lie on one of the longitude lines of the quadrilateral.
To determine the longitude edge closest to $p$,
we calculate the center longitude of $Q$ and rotate it to align with the meridian.
We rotate $p$ by the same amount.
If the longitude of $p$ is now positive, the west longitude-edge of $Q$ is closer to $p$, otherwise the east one.\footnote{%
  This is possible because we split the Earth at the antimeridian.
  Therefore, the western longitude of $Q$ is always smaller than the eastern one.}
Having determined the closest longitude edge,
we can now calculate point $s$ using linear algebra in Euclidean space as described in \cref{sec:app:distance}.
If the latitude of $s$ is between the top and bottom latitude of $Q$,
 $s$ is the point with the shortest distance to $p$ in $Q$ (case \ref{distancequadr:case3}).
Otherwise one of the corners is the point with the shortest distance to $p$ (case \ref{distancequadr:case4}).

Given these primitives, insert and query operations on space-partitioning trees for spherical surfaces are
identical to the ones for Euclidean space.

\section{Implementation}\label{sec:impl}

We implement our new sweep-plane algorithm in the \textsc{mountains} C++ framework by \kirmseAuth \cite{kirmse2017calculating}.\footnote{%
  \url{https://github.com/akirmse/mountains}}
The framework provides essential functionalities for the work with tiled DEM data, such as data loading and conversion,
peak discovery and distance computations.
Our code is available on Github.\footnote{%
\url{\GitUrl}}
In the following we provide details our implementation.

\subparagraph*{Data Structure.}

We implement a fully dynamic $k$-D-Tree, that supports insert and remove operations in \Ologn worst case time
and nearest neighbor searches in \Ologn expected time \cite{friedman1977algorithm} using the predicates described in the previous section.\footnote{%
We also adapted a Quadtree using our predicates, which was however outperformed by the $k$-D-Tree in our experiments.}
Points are stored in the leaves of the tree, which have a fixed capacity of $C$ points.
On exceeding capacity $C$, a leaf is split according to a center-split policy along the longer side of the quadrilateral.
The first points inserted into the tree often belong to the highest peak in a tile and are thus close to each other.
In order to prevent the tree from degenerating, we pre-build the first $k$ levels in a Quadtree-like manner.\footnote{%
Our experiments show $k = 4$ to be a good choice.}
To improve cache-efficiency, tree nodes are allocated in blocks and are re-used after deletion.

Another data structure used is the \tpMap,
which we implement as  an internal  memory hash map with the latitude and longitude of a tile as key and a list of peaks as value.

\subparagraph*{Algorithm.}

Given the search area $A$, all contained DEM tiles can be processed in parallel.
We use a work queue and thread pool to distribute the computations among the available processing elements.
The \tpMap is initialized in advance with all tiles.
To sort the events of a pass, we use the efficient sorting algorithm ips4o of Axtmann \etal \cite{ips4o}.

\subparagraph*{Bounding Pass.}
We use the peak detection algorithm of \kirmseAuth \cite[Sec. 2.2]{kirmse2017calculating} to find all peaks within a tile.
It considers all points to be peaks, that have eight neighboring points of lower or equal elevation.
If a peak consists of several equally high sample points, only a single one is added to the set of peaks.\footnote{%
The algorithm by \kirmseAuth always chooses the north-west corner of a peak.}
Since only an upper bound is calculated during the bounding pass,
we down-sample the resolution of the DEM after peak detection.
Since all peaks are within the tile which is processed,
distances are rather small and can be approximated using planar Euclidean geometry during the nearest-neighbor search.
The upper bound is computed using the spherical distance function according to \cref{eq:app:distanceCircle}.
If the determined upper bound on the isolation of a peak is below a threshold $I_\text{min}$
we discard the peak as insignificant.
Peaks with an upper bound above $I_\text{min}$ are added to the \tpMap for processing in the finalization pass as described in \cref{sec:algo:scalable}.

\subparagraph*{High-Point Pass.}
While processing the tiles in the bounding pass, 
we build a geometric search tree $T$ that partitions the entire search area down to the tile level.
Internal nodes of $T$ save the highest elevation in its sub-tree.
After all tiles have been processed,
we can use this information to efficiently determine upper bounds on the isolation of the high-points of each tile.
For each high-point $h$, we find the closest tile containing a point of higher elevation than $h$ in search tree $T$.
The maximum distance between $h$ and any point within the found tile serves as an upper bound on $h$'s isolation.
Given this upper bound we can add $h$ to all tiles containing potential ILPs in the \tpMap.
This approach is trivially parallelizable over the number of high-points in the search area.

\subparagraph*{Finalization Pass.}
In this pass we use the full resolution of the input DEM.
For each tile, the peaks processed in this pass are the ones that are assigned to it in the \tpMap.
We use ellipsoid distance computations according to \cref{eq:app:distanceEllipsoid}. %

\FloatBarrier
\section{Evaluation}\label{sec:Evaluation}

In this section we evaluate our novel sweep-plane algorithm to calculate the isolation of mountain peaks, which we named \ourAlg. 
We evaluate it with regard to runtime behavior and solution quality and compare it against \kirmseAlg from \kirmseAuth \cite{kirmse2017calculating}.

\subparagraph*{Experimental Setup.}

All benchmarks are conducted on a machine with an AMD EPYC Rome 7702P with 64 cores and 1024\,GB of main memory.
The DEM data is stored on a Intel P4510 2\,TB NVMe SSD.
We use the \qty{3}{\arcsecond} data set from viewfinderpanoramas \cite{viewfinderpanoramas} with worldwide coverage.
To build smaller test instances from the worldwide data set,
we choose a random starting tile and add neighboring tiles in a spiraling manner around it until the desired number of tiles is reached.
For each instance size, we generate several instances to cover a wide range of terrain,
refer to \cref{tab:app:testDataset} for details.
Additionally, we use the \qty{3}{\arcsecond} and \qty{1}{\arcsecond} DEM for the USA, Canada and most of Europe from viewfinderpanoramas \cite{viewfinderpanoramas} to study the scaling behavior for higher-resolution DEMs. 
This data set  corresponds to  \num{4294} tiles or roughly \qty{16}{\percent} of the total test data set.
We will call this data set NA-EU.

For all benchmarks we use an isolation threshold $I_\text{min}$ of \qty{1}{\kilo\metre} and report the mean runtime of $5$ runs. 
I/O costs are not part of the reported figures as we use the \textsc{mountains} framework \cite{kirmse2017calculating} 
for them without an attempt at optimization.
They are about the same time as computation for \qty{3}{\arcsecond} DEMs and about \qty{10}{\percent} of computation time
for \qty{1}{\arcsecond} ones
and thus could be overlapped with the computation in an optimized framework.

\subsection{Runtime and Scaling Behavior.}

The runtime of the algorithms depends on the number of sample points in the DEM.
These can either increase due to a larger search area or a higher-resolution DEM.
Another factor is the number of processed peaks, 
since every peak starts a local search in \kirmseAlg and a nearest neighbor query in \ourAlg.
A larger search area increases the number of tiles and the number of peaks, 
whereas higher-resolution data increases mostly the number of points per tile.
High-resolution DEMs can contain more peaks than lower-resolution ones due to the more truthful representation of the terrain,
however these are predominantly low-isolation peaks and are filtered out in the first pass.
We study both effects in our experiments by using different resolution DEMs as well as increasing search areas.

\begin{figure}[tbp]
	\centering
	\begin{subfigure}{\textwidth}
		\centering
		\includegraphics[width=\textwidth]{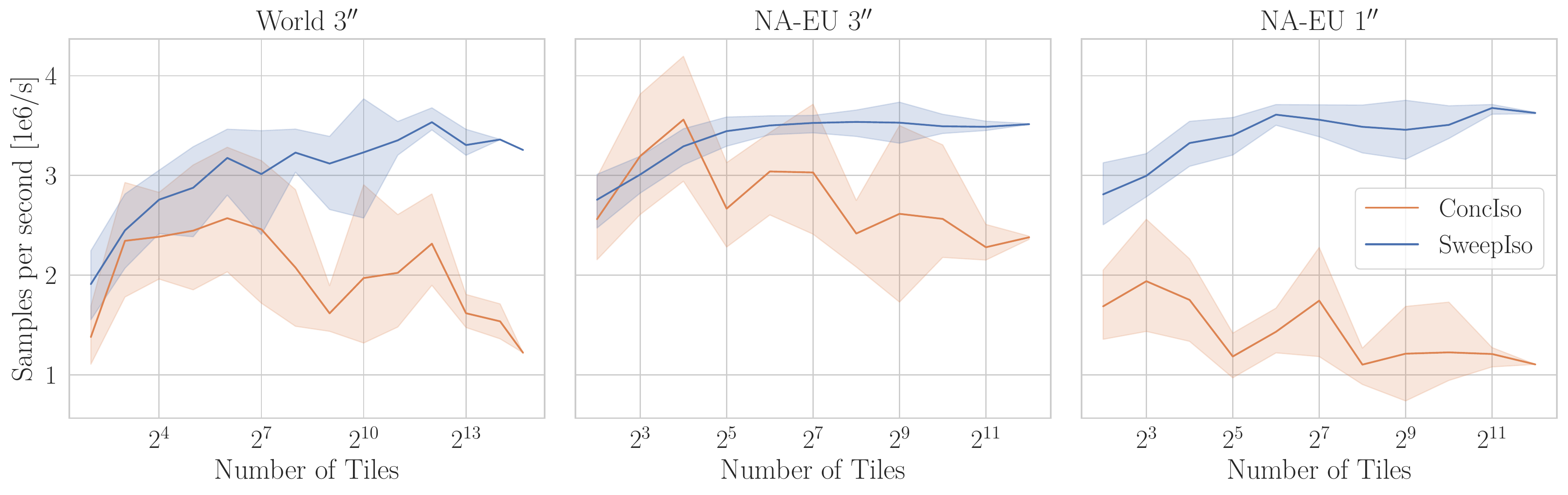}
		\caption{Throughput in sample points per second over number of tiles.}
		\label{fig:runtime:pts}
	\end{subfigure}
	\begin{subfigure}{\textwidth}
		\centering
		\includegraphics[width=\textwidth]{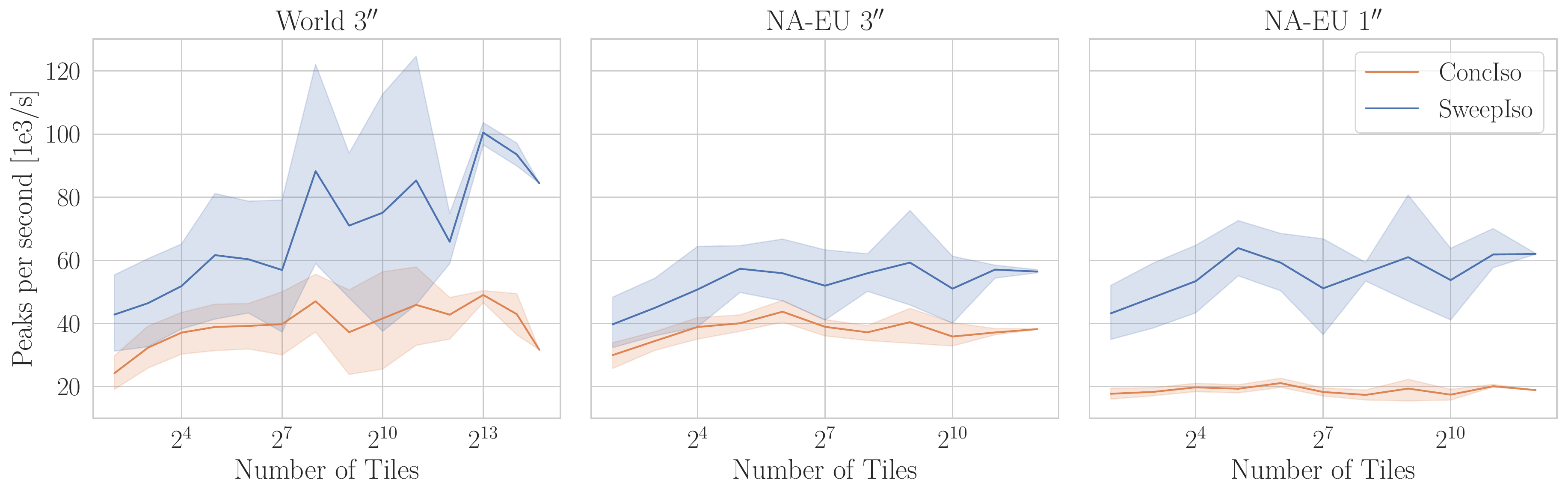}
		\caption{Throughput in processed peaks per second over number of tiles.}
		\label{fig:runtime:pks}
	\end{subfigure}
	\caption{Single-threaded runtime comparison of \ourAlg and \kirmseAlg.}
	\label{fig:runtime}
\end{figure}

\begin{figure}[tbp]
	\centering
	\begin{subfigure}{.50\textwidth}
		\centering
		\includegraphics[width=\textwidth]{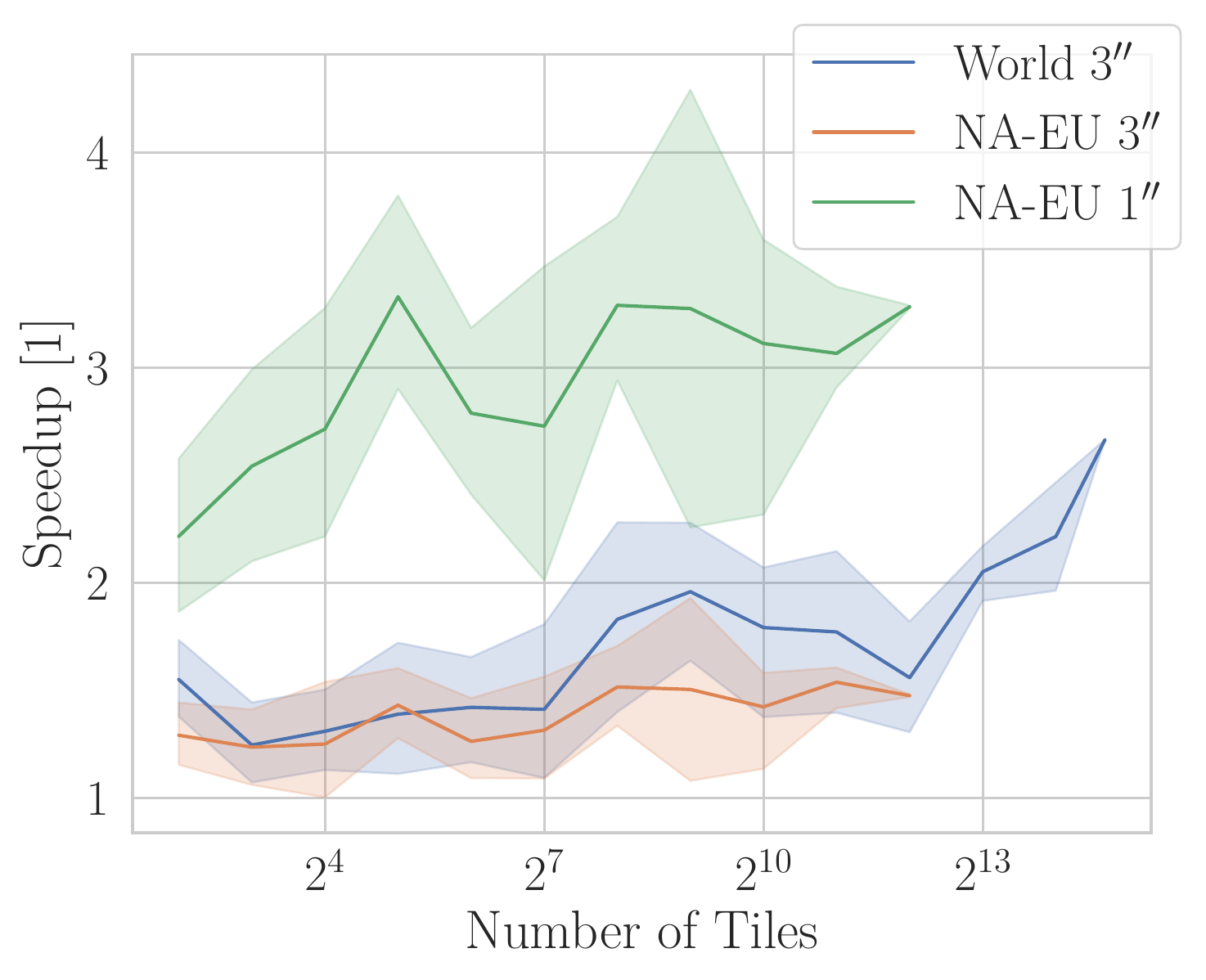}
		\caption{Single-threaded speedup.}
		\label{fig:speedup:sc}
	\end{subfigure}\hfill
	\begin{subfigure}{.40\textwidth}
		\centering
		\includegraphics[width=\textwidth]{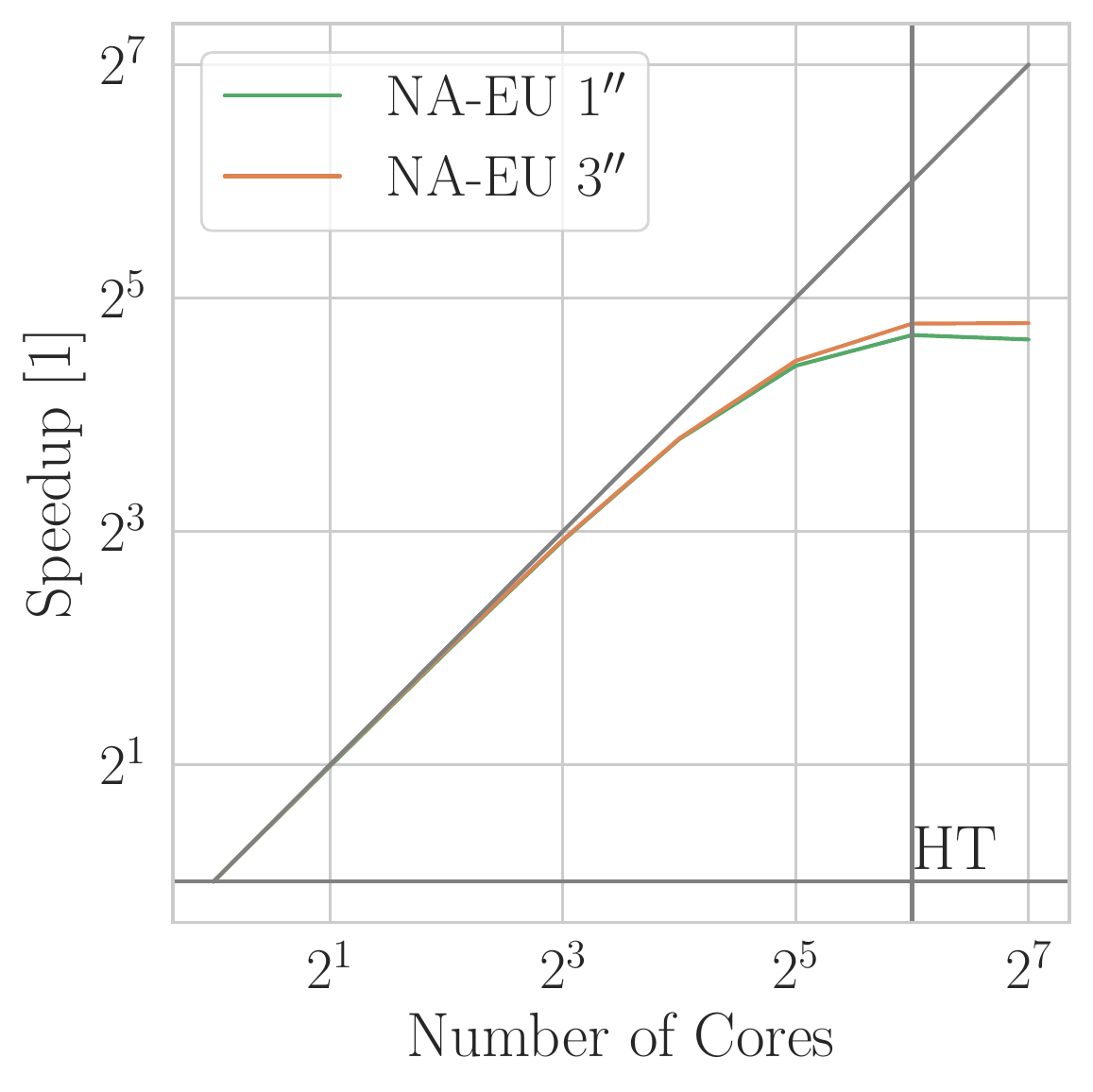}
		\caption{Multi-threaded relative speedup.}
		\label{fig:speedup:mc}
	\end{subfigure}
	\caption{Single-threaded speedup of \ourAlg over \kirmseAlg and relative speedup of \ourAlg for multi-threaded execution.}
\end{figure}

\ourAlg exhibits a nearly constant throughput of sample points per second with increasing search area,
while \kirmseAlg's throughput degrades -- refer to \cref{fig:runtime:pts}.
\Cref{fig:speedup:sc} shows that \ourAlg outperforms \kirmseAlg by a factor of \numrange{2}{3} in terms of sample point throughput.
For instance, we reduce the time required to compute the isolation of every peak on Earth from \qty{9}{\hour} down to \qty{3.5}{\hour}.
However, \kirmseAlg's runtime scales significantly better than its \OnSq worst case runtime bound would suggest.
This is because most peaks have a relatively low isolation.
In fact, \qty{99.996}{\percent} of discovered peaks on the world data set have an isolation below \qty{50}{\kilo\metre} and more than \qty{99}{\percent} below \qty{10}{\kilo\metre}.
Since one tile covers on average an area of $\qty{70}{\kilo\metre} \times \qty{111}{\kilo\metre}$,
the nearest higher point is most of the time within the same tile as the peak,
where fast approximations for the distance calculation are used.
Nevertheless, for high-resolution DEMs the computation cost per peak is significantly higher for \kirmseAlg than for \ourAlg,
refer to \cref{fig:runtime:pks}.
\Cref{fig:app:runtime} in the appendix shows detailed runtime data for all test instances.

\begin{figure}[tbp]
	\centering
	\includegraphics[width=0.7\textwidth]{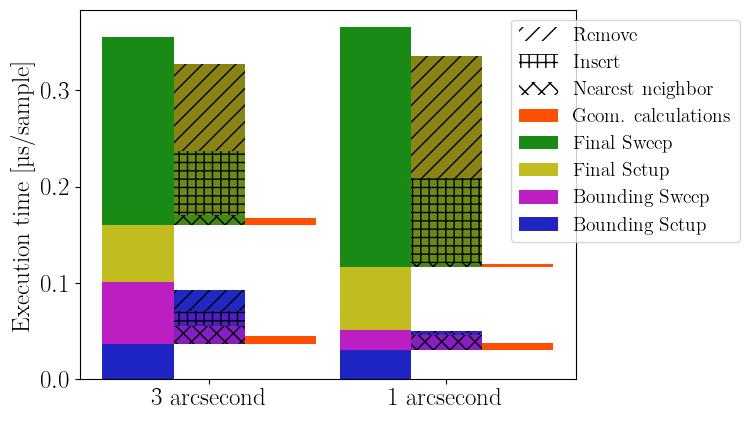}
	\caption{Runtime composition of the bounding and finalization pass for  \qty{3}{\arcsecond} and \qty{1}{\arcsecond} data.
		The sweep-plane operations of these passes are further subdivided into remove, insert, nearest neighbor search operations as well as the time for distance calculations.}
	\label{fig:operationtimes}
\end{figure}

\Cref{fig:operationtimes} shows the runtime composition of the bounding and the finalization pass. 
As the DEM resolution is reduced in the bounding pass, the finalization pass requires the majority of the runtime,
especially for higher-resolution inputs.
Geometric computations only make a small fraction of the runtime.
The high-point pass requires just \qty{0.001}{\percent} of the total execution time
and is therefore omitted in the figure.

\Cref{fig:speedup:mc} shows the results of our multi-threaded runtime experiments with the NA-EU data set.
\ourAlg scales well with physical cores, but does not benefit from hyper-threading (HT).
For \num{64} cores it reaches a speedup of \num{25}.
We verified the scaling behavior of the algorithm for the entire Earth
and where able to confirm a speedup of \num{25} on \num{64} cores.
This corresponds to a runtime of \qty{8}{\minute} to calculate the isolation of every peak on Earth.
We were not able to execute \kirmseAlg with multiple threads due to issues in the implementation.

\subsection{Solution Quality}\label{sec:Evaluation:Accuracy}

\begin{figure}[tbp]
		\centering
	\includegraphics[width=.45\textwidth]{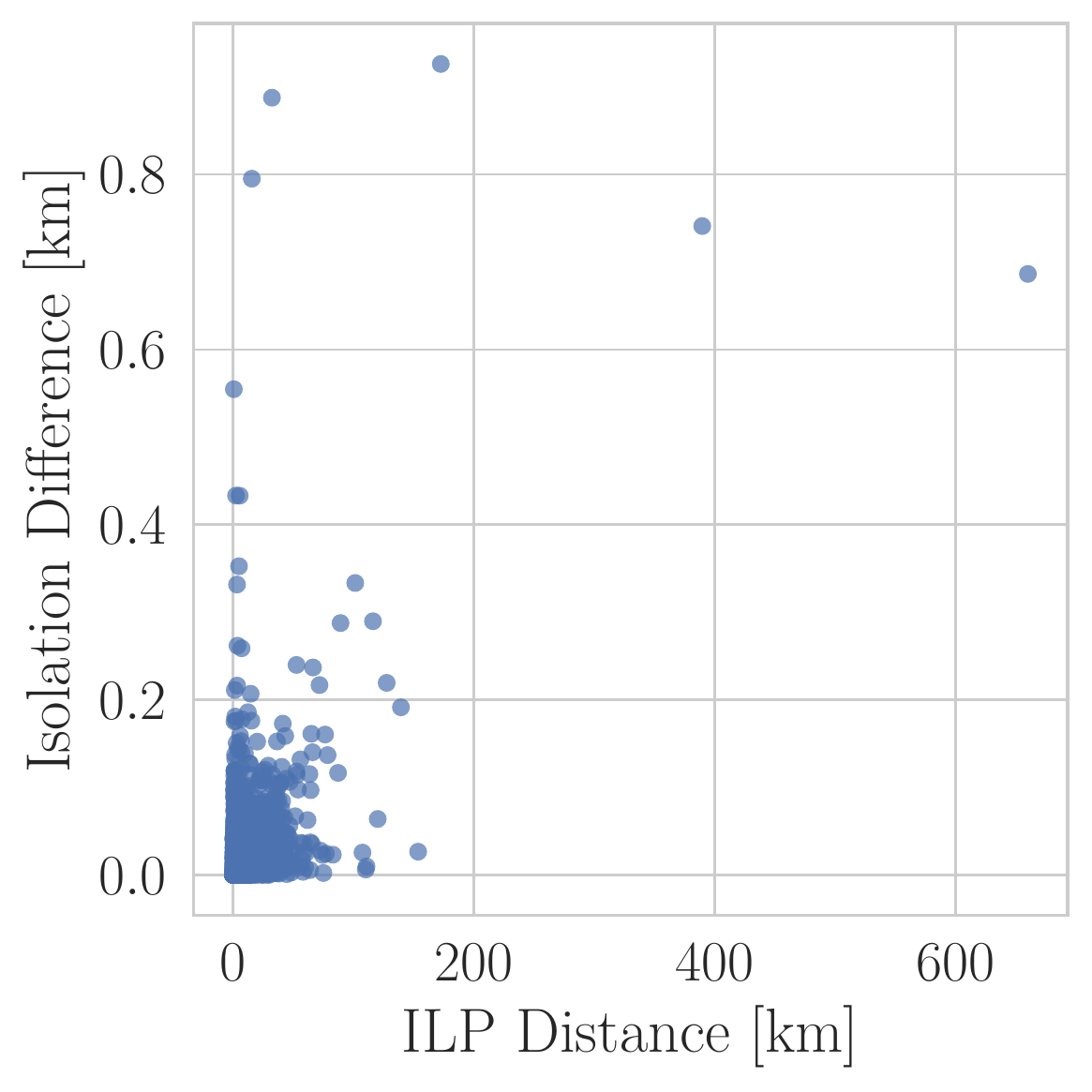}
		\caption{Comparison of difference in isolation between \ourAlg and \kirmseAlg and distance between found ILPs.}
		\label{fig:isoDiff}
\end{figure}

In principle, at any particular time, the isolation and witnessing ILP of a peak are a well defined. 
However the actually computed values depend on imprecisions in both the data and the used algorithms.

Due to the design of \ourAlg, more expensive distance approximations can be used to find the ILP as in \kirmseAlg.
This results in closer ILPs being found as shown in \cref{fig:isoDiff}, 
which displays the distribution of the difference in isolation and the distance between the 
found ILPs between \ourAlg and \kirmseAlg, using the same data.
Even if the isolation values between the two algorithms do not vary greatly, 
the distances between the ILPs do. 
For example for the Cerro Gordo summit in Mexico 
both algorithms find ILPs that are more than \qty{600}{\kilo\metre} apart
while \ourAlg's ILP is merely \qty{.8}{\kilo\meter} closer to the peak.

To further evaluate the results, we used the collection of peaks with more than \qty{300}{\kilo\meter} 
isolation from the website \url{peakbagger.com} \cite{peakbagger}. 
The comparison showed that for about \qty{75}{\percent} of peaks
in this list the isolation deviation is below \qty{2}{\kilo\meter}. 
In a DEM, a sample point corresponds to the average elevation of the area it represents. 
This often leads to an underestimation of a peak's elevation, sometimes significantly \cite{kirmse2017calculating}.
For instance, according to our DEM data, Galdhøpiggen (the highest point of Scandinavia)
is \qty{11}{\meter} lower than Glittertind.
This causes a significant change in isolation of both mountains.
\cref{tab:app:isolation-diff} lists the five summits with the biggest differences in isolation between peakbagger and our calculation.

\section{Conclusion and Future Work}\label{sec:Conclusion}
We have presented
\ourAlg, a scalable and efficient
algorithm to compute the isolation of peaks.
\ourAlg considerably outperforms the previous more
brute-force state-of-the-art approach. The performance
gains also enable more accurate distance
calculations at decisive places resulting in higher accuracy.  \ourAlg is able to
process the entire earth for currently publicly
available data within minutes. This is relevant
as it indicates that \ourAlg can also handle
higher-resolution data that is available
commercially or that will be available in the
future. \ourAlg's two-level semi-external sweeping
architecture may also be an interesting design pattern for
other computations on massive DEM data.
Furthermore, \ourAlg could serve as a
benchmark for dynamic nearest neighbor search data
structures. 

\subparagraph{Future Work.}
From an application perspective, it
would be interesting to compute not only
isolations for a given, necessarily imprecise data
set but to compute confidence bounds that take
into account error margins in the input data.
This would be possible at a moderate increase in
cost. Peaks could be replaced by enclosing boxes/circles while vertical errors
could be handled
by having ``may-be-there'' and
``must-be-there'' insertion events and sweep-plane
data structures. Geographically most interesting
would be those isolations that change a lot
depending on how high exactly particular pairs of
peaks are.  Those pairs could then be valuable
targets for additional data cleaning or new
measurements.

For algorithm engineering, it would be interesting
to close the gap between theory and practice with
respect of nearest-neighbor data structures.  We
have reasonable empirical performance of simple
data structures like $k$-D trees but
no well-fitting performance guarantees applicable to
\ourAlg.  For example, one could look for a more
general characterization of inputs where cover
trees \cite{covertree} work provably well.

\newpage
\bibliographystyle{plainurl}
\bibliography{thesis}

\newpage
\appendix

\FloatBarrier

\section{Basic Mathematical Concepts} \label{sec:app:distance}

\subsection{Spherical Distance Calculations}
Since planets are often approximated by spheres, a lot of calculations on the sphere's surface take place \cite{WGS}. 
The geographic coordinate system is used, where points are given in latitude and longitude. 
The latitude is given as an angle between $0^\circ$ and $90^\circ$ from the equator in the north or south direction and the longitude from $0^\circ$ to $180^\circ$ east, west in respect to the meridian. We will also use the notation where the south and west values are given in negative degrees. 
So the point $(20^\circ S, 90^\circ W)$ can also be written as $(-20^\circ,-90^\circ)$.

On the sphere, surface great circles correspond to straight lines in Euclidean space. 
Thus, there are always at least two great circle segments as straight lines between two points on the sphere surface. 
The smallest great circle segment between two points is called the geodesic. 
If the points are at polar ends of the sphere there exists an infinite number of geodesics between them.

The following formula is used to calculate the length of the geodesic for points $(\lambda_1, \phi_1)$, $(\lambda_2, \phi_2)$ and radius $R$.

\begin{align} \label{eq:app:distanceCircle}
  a               & = \sin^2\left(\frac{\phi_2 - \phi_1}{2}\right) + \sin^2 \left(\frac{\lambda_1 - \lambda_2}{2}\right) \cos(\lambda_1) \cos(\lambda_2) \\
  \text{distance} & = R \tan^{-1}\left( \frac{\sqrt{a}}{\sqrt{(1-a)}}\right) \nonumber
\end{align}

The earth and most other planets are not perfect spheres.
An ellipsoid is a better approximation, yielding the following distance formula:
\begin{align} \label{eq:app:distanceEllipsoid}
  \text{d}\lambda & := (\lambda_2 - \lambda_1)/2 \quad \text{d}\phi: = (\phi_2 - \phi_1)/2                                                                                                            \\
  \Lambda         & = (\lambda_2 + \lambda_1) / 2 \nonumber                                                                                                                                           \\
  s               & = \sin^2 \text{d}\phi \cdot \cos^2 \text{d}\lambda + \cos^2\Lambda \cdot \sin^2 \text{d}\lambda \nonumber                                                                         \\
  c               & = \cos^2\text{d}\phi \cdot \cos^2 \text{d}\lambda + \sin^2\Lambda \cdot \sin^2 \text{d}\lambda \nonumber                                                                          \\
  w               & = \tan^{-1}(\sqrt{s} / \sqrt{c} ) \nonumber                                                                                                                                       \\
  r               & = \frac{\sqrt{sc}}{w} \nonumber                                                                                                                                                   \\
  \text{distance} & = 2aw\left( 1 + f \frac{3r - 1}{2c}\sin^2 \Lambda \cdot \cos^2 \text{d}\phi - f \frac{3 r + 1}{2c} \sin^2 \Lambda \cdot \sin^2\text{d}\phi \cdot \cos^2 \Lambda \right) \nonumber
\end{align}
$r$ and $f$ are the equatorial radius and flattening of the World Geodetic System (WGS) \cite{WGS}.

\subsection{Quadrilateral Computations}
Given a quadrilateral $Q$ defined by its north-east and south-west corner and a query point $p \not\in Q$,
we need to compute the closest point $s \in Q$ to $p$.
First, the longitude-edge of $Q$ which is closer to $p$ is determined as described in \cref{sec:SphereTree}.
Point $s$ must lie on this edge and can be computed using linear algebra.
For this point $p$ and the edge-points defining the longitude edge, $l_S$ and $l_N$, are transformed according to 
\begin{equation}\label{eq:app:toCartesian}
	x := \cos(\operatorname{lat}(x)) \cos(\operatorname{long}(x)) \qquad
	y := \cos(\operatorname{lat}(x)) \sin(\operatorname{long}(x)) \qquad
	z := \sin(\operatorname{lat}(x))
\end{equation}
This transformation assumes Earth to be a sphere.
For a more accurate transformation assuming Earth to be an ellipsoid refer to \cite{WGS}.
Now, the closest point $s$ to $p$ can be determined according to
\begin{equation}\label{eq:app:findNearestPointOnGCSegment}
	A := l_S \times l_N \qquad
	B := p \times A \qquad
	S := A \times B
\end{equation}
Note, $S$ is normalized, because \cref{eq:app:toCartesian} produces normalized vectors 
and \cref{eq:app:findNearestPointOnGCSegment} uses only cross products.
The idea behind these formulae is the following:
$A$ is the plane of the great circle defined by $l_N$ and $l_S$.
$B$ is the plane of the great circle that is perpendicular to $A$ and goes through the point $p$.
Lastly, the intersection between these two great circles is calculated ($S$),
yielding the geodesic between $s$ and $p$, that is perpendicular to the longitude circle and thus the shortest possible one.
$S$ is now transferred back to longitude and latitude using
\begin{equation*}
	\operatorname{lat}(s) := \arcsin(z) \qquad \operatorname{long}(s) := \arctan_2(x,y)
\end{equation*}
This yields point $s$ on the longitude-edge of $Q$ that is closest to $p$.
If $s$ lies outside $Q$'s top or bottom latitude, the respective corner of $Q$ is the closest point to $p$.

\begin{figure}[tb]
  \centering
  \begin{subfigure}[t]{.3\textwidth}
    \centering
    \includegraphics[width=\textwidth]{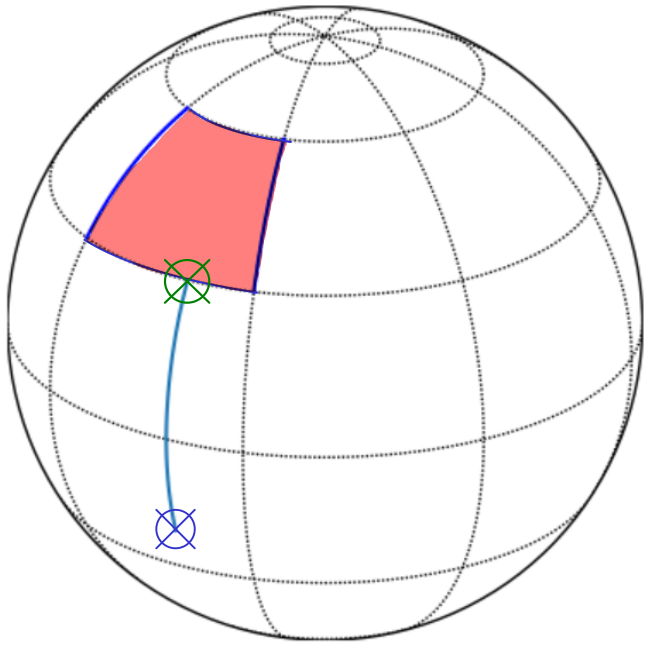}
    \caption{Case \ref{distancequadr:case2}.}
  \end{subfigure}\hspace{.04\textwidth}
  \begin{subfigure}[t]{.3\textwidth}
    \centering
    \includegraphics[width=\textwidth]{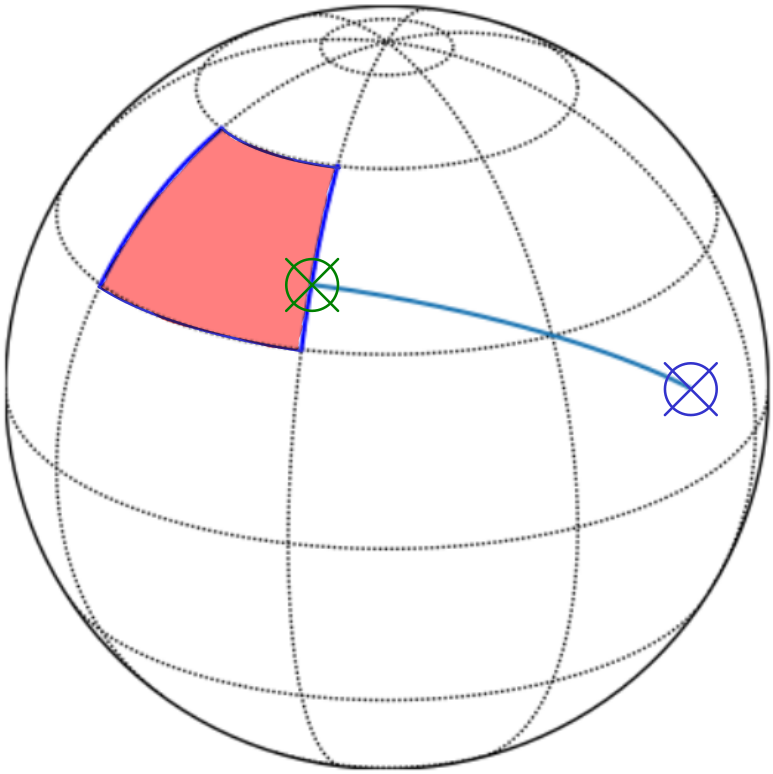}
    \caption{Case \ref{distancequadr:case3}.}
  \end{subfigure}\hspace{.04\textwidth}
  \begin{subfigure}[t]{.3\textwidth}
    \centering
    \includegraphics[width=\textwidth]{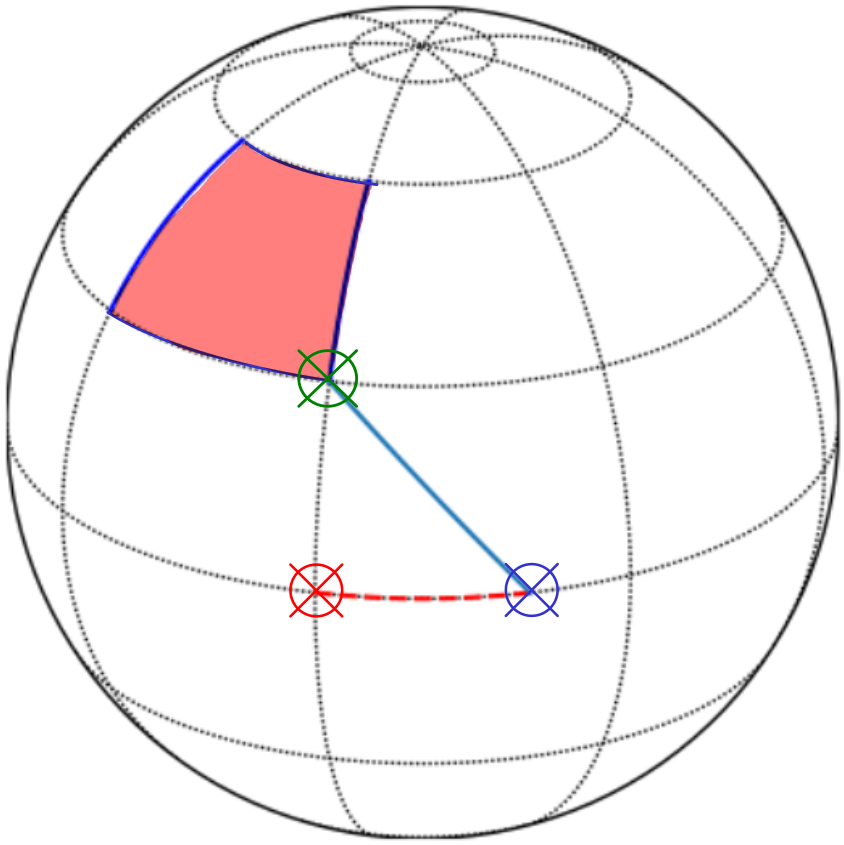}
    \caption{Case \ref{distancequadr:case4}.}
  \end{subfigure}
  \caption{Min-distance between quadrilateral and point examples.}
  \label{fig:app:distance-quadrilateral}
\end{figure}

\FloatBarrier

\FloatBarrier
\newpage
\section{Evaluation}\label{sec:app:eval}

\begin{figure}[tbh]
	\centering
	\begin{subfigure}[t]{.3\textwidth}
		\centering
		\includegraphics[width=\textwidth]{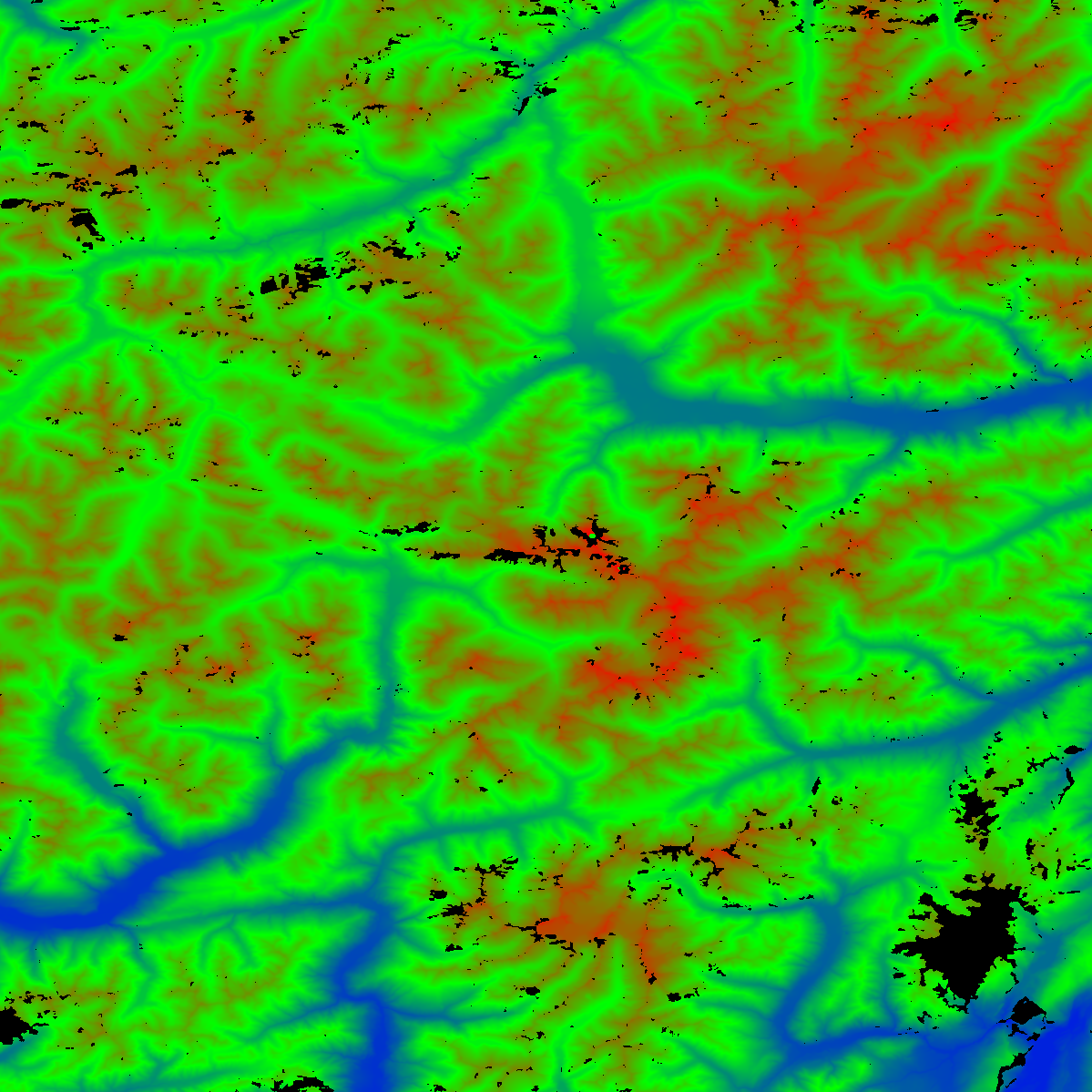}
	\end{subfigure}\hspace{.1\textwidth}
	\begin{subfigure}[t]{.3\textwidth}
		\centering
		\includegraphics[width=\textwidth]{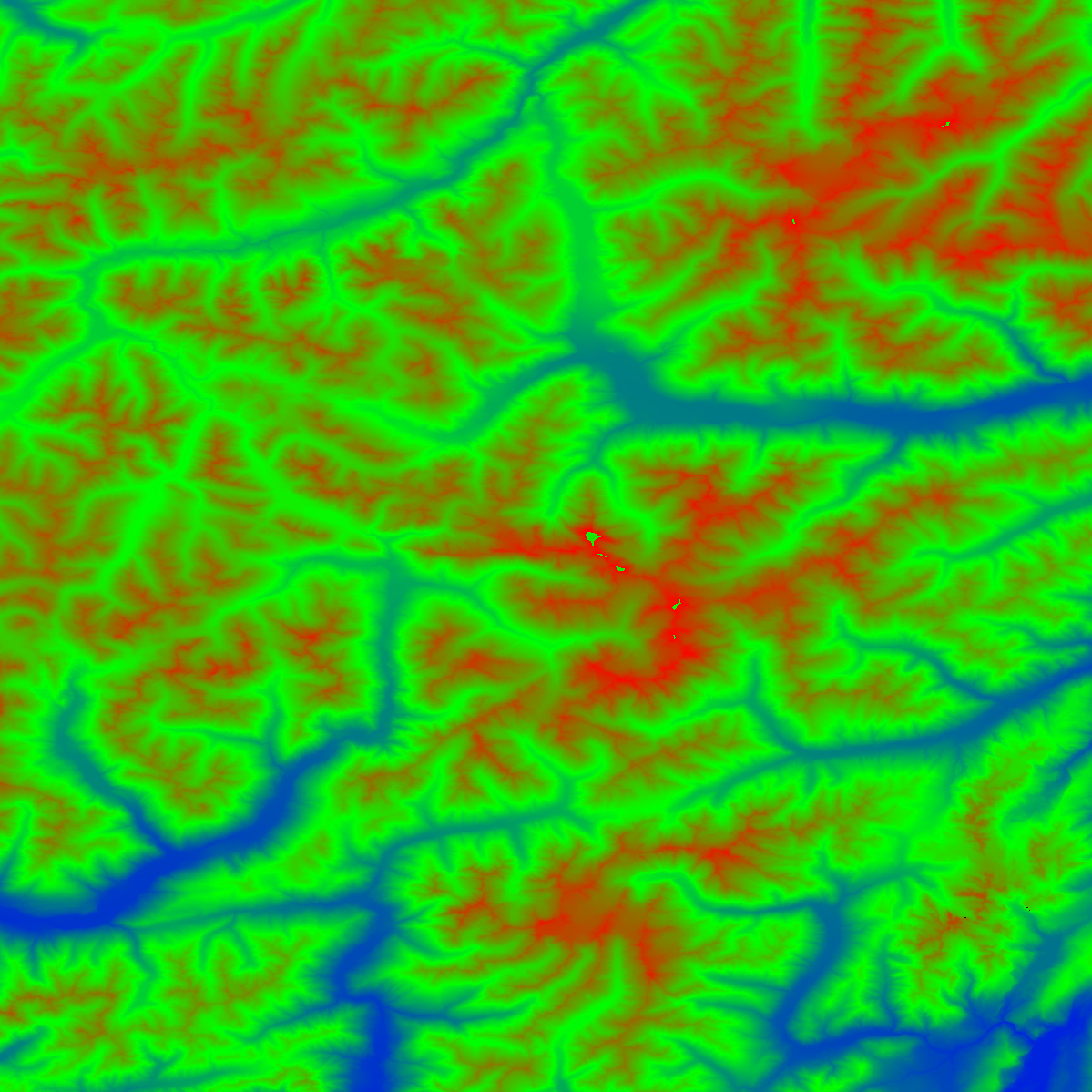}
	\end{subfigure}
	\caption{Comparison tile $46^\circ N$ $10^\circ E$ (\"Otztal and Ortler Alps). Left raw data from NASA's SRTM with voids (black), right enhanced data by viewfindpanoramas.org.}
	\label{fig:app:vergleich-srtm}
\end{figure}

\begin{table}[tbh]
	\centering
			\caption{Test instances used for worldwide and NA-EU execution time testing.}
    	   \label{tab:app:testDataset}
	\begin{subtable}[t]{0.48\textwidth}
			\centering
         \caption{Worldwide data set.}
		\begin{tabular}{rr}
        \toprule
			Tile count & Random samples  \\
			\midrule                               \\
			\num{4}     & \num{50}    \\
			\num{8}     & \num{40}    \\
			\num{16}    & \num{32}   \\
			\num{32}    & \num{24}   \\
			\num{64}    & \num{18}   \\
			\num{128}   & \num{12}  \\
			\num{256}   & \num{8}    \\
			\num{512}   & \num{4}    \\
			\num{1024}  & \num{4}   \\
			\num{2048}  & \num{4}   \\
			\num{4096}  & \num{3}   \\
			\num{8192}  & \num{3}   \\
			\num{16384} & \num{2}  \\
			\num{26095} & \num{2}  \\
            \bottomrule
		\end{tabular}
        \label{tab:app:testDataset:world}
	\end{subtable}
		\begin{subtable}[t]{0.48\textwidth}
		\centering
                 \caption{NA-EU data set.}
        		\begin{tabular}{rr}
                \toprule
        			Tile count & Random samples  \\
        			\midrule                               \\
					\num{4}     & \num{50}    \\
					\num{8}     & \num{40}    \\
					\num{16}    & \num{32}   \\
					\num{32}    & \num{24}   \\
					\num{64}    & \num{18}   \\
					\num{128}   & \num{12}  \\
					\num{256}   & \num{8}    \\
					\num{512}   & \num{4}    \\
					\num{1024}  & \num{4}   \\
					\num{2048}  & \num{4}   \\
					\num{4096}  & \num{3}   \\
                    \bottomrule
        		\end{tabular}
        \label{tab:app:testDataset:us}
	\end{subtable}
\end{table}

\renewcommand{\cellalign}{tl}
\renewcommand\theadalign{bc}
\renewcommand\theadfont{\bfseries}
\begin{table}[tbh]
    \caption{Biggest isolation differences between discovered and peakbagger (PB) data.}
    \label{tab:app:isolation-diff}
	    \begin{tabularx}{\textwidth}{lrrrX}%
	    	\toprule \\
		\thead[bl]{Mountain} & \thead{PB \\ Rank} & \thead{PB \\ Isolation} & \thead{\ourAlg \\ Isolation} & \thead[bl]{Notes} \\
		\hline \\
		\begin{filecontents*}{data/diff-pb-notes.csv}
peakLat,peakLng,name,pbrank,old,new,notes
61.636429,8.312562,\makecell{Galdhøpiggen \\ (Norway)},47,1568.3,12.9,\makecell{\textit{Galdhøpiggen} \\ DEM elev.: \qty{2455}{\meter} -- real elev.: \qty{2469}{\meter} \cite{peakvisor} \\ \textit{Glittertind} \\ DEM elev.: \qty{2466}{\meter} -- real elev.: \qty{2457}{\meter} \cite{peakvisor} \\ Isolation \qty{1563.5}{\kilo\meter}}
-84.332214,166.431091,\makecell{Mt. Kirkpatrick \\ (Antarctica)},45,1585.2,53.5,\makecell{found ILP at -83.8967;168.3733 \\ DEM elev.: \qty{4416}{\meter} -- real elev.: \qty{4528}{\meter} \cite{peakvisor} \\ ILP elev.: \qty{4416}{\meter}}
-69.79437,-64.495881,\makecell{Mt. Hope \\ (Antarctica)},75,1113.5,203.4,No data found in other sources
-75.028428,123.554706,\makecell{Dome Charlie \\ (Antarctica)},92,971.8,248.6,\makecell{found ILP at -76.3783;116.3708 \\ DEM elev.: \qty{3265}{\meter} -- real elev.: \qty{3233}{\meter} \cite{antarcticaStationCat} \\ ILP elev. \qty{3266}{\meter}}
-13.6181,-172.4866,\makecell{Mauga Silisili \\ (Samoa)},23,2245.1,2502.8, \makecell{DEM elev.: \qty{1854}{\meter} -- real elev.:\qty{1863}{\meter} \cite{peakvisor} \\ PB ILP elev.: \qty{1879}{\meter} \\ found ILP elev.: \qty{1836}{\meter}}
-12.20836,96.895046,\makecell{Cocos Islands \\ High Point},94,961.4,947.4,Found ILP at Enggano Island
\end{filecontents*}
\csvreader[head to column names]{data/diff-pb-notes.csv}{}%
		{\name & \pbrank & \old & \new & \notes \\}%
		\\ \bottomrule
	\end{tabularx}
\end{table}

\begin{figure}[tbh]
	\centering
	\begin{subfigure}{\textwidth}
		\centering
		\includegraphics[width=\textwidth]{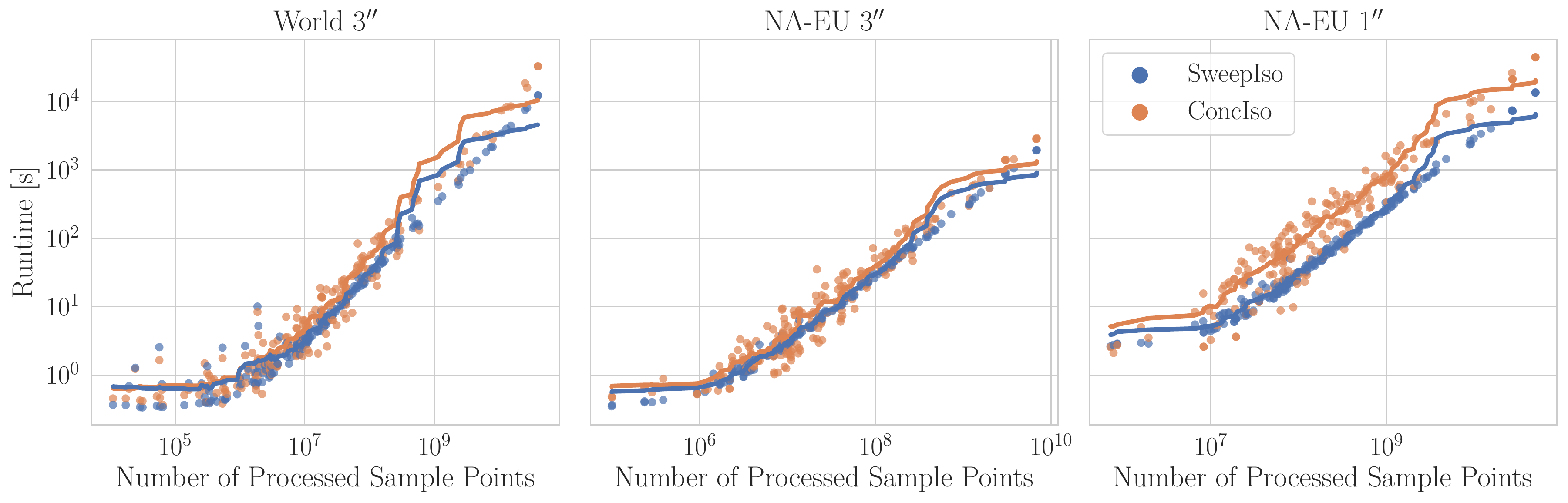}
		\caption{Execution time over processed sample points.}
	\end{subfigure}
	\begin{subfigure}{\textwidth}
		\centering
		\includegraphics[width=\textwidth]{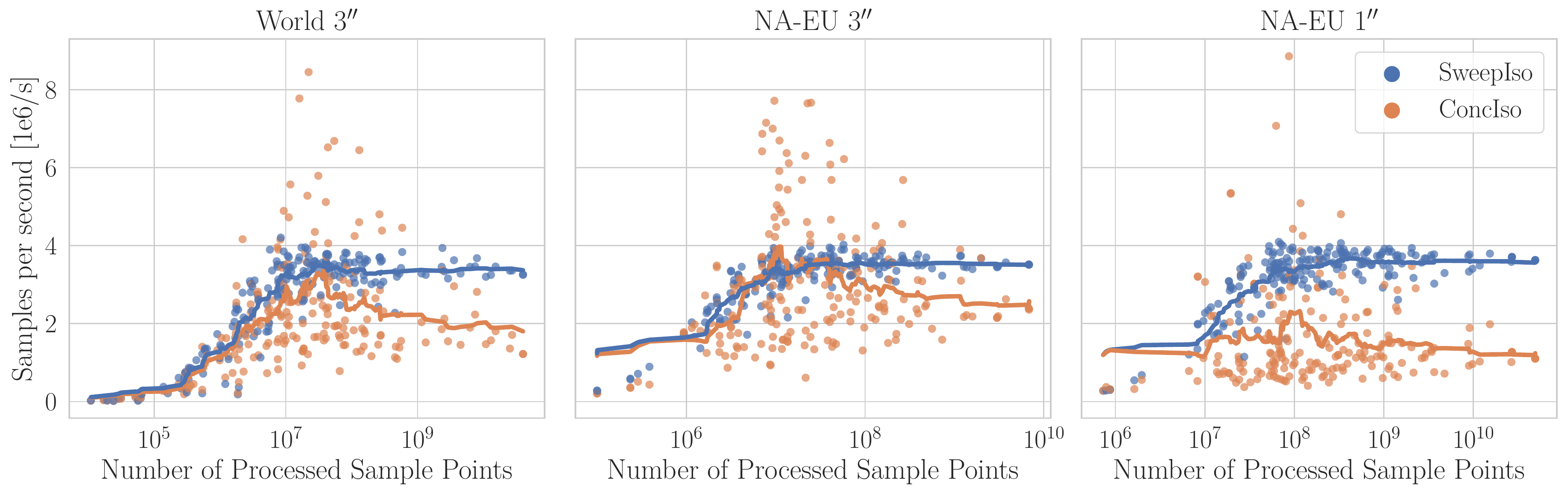}
		\caption{Throughput in points per second over processed sample points.}
	\end{subfigure}
	\begin{subfigure}{\textwidth}
		\centering
		\includegraphics[width=\textwidth]{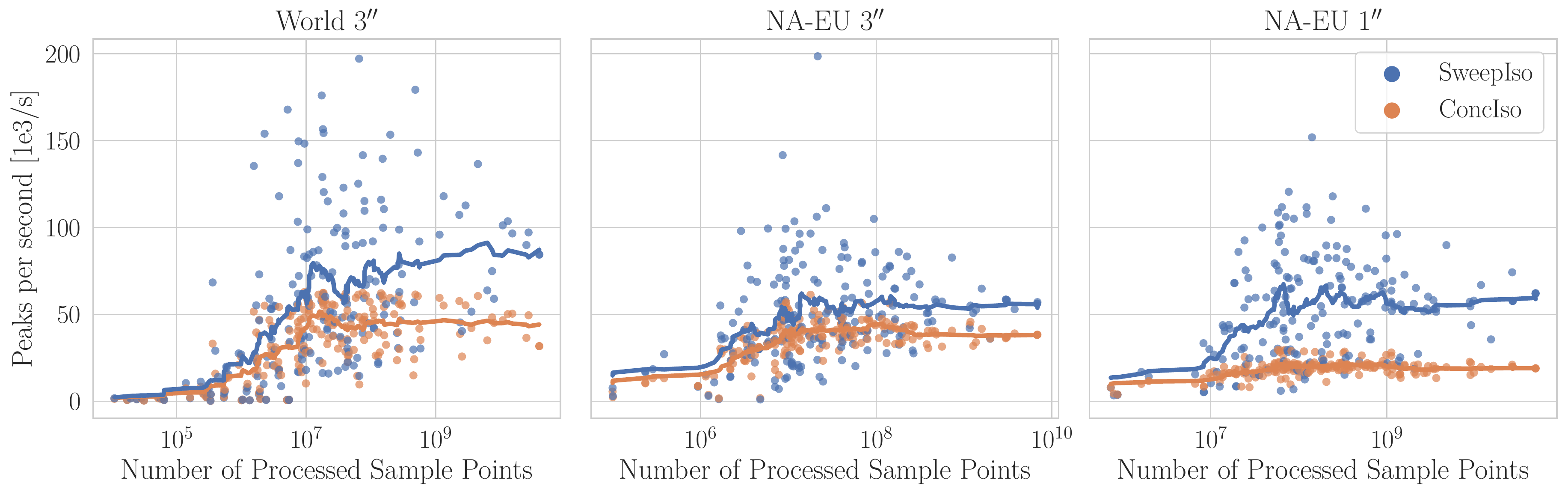}
		\caption{Throughput in peaks per second over processed sample points.}
	\end{subfigure}
	\begin{subfigure}{\textwidth}
		\centering
		\includegraphics[width=\textwidth]{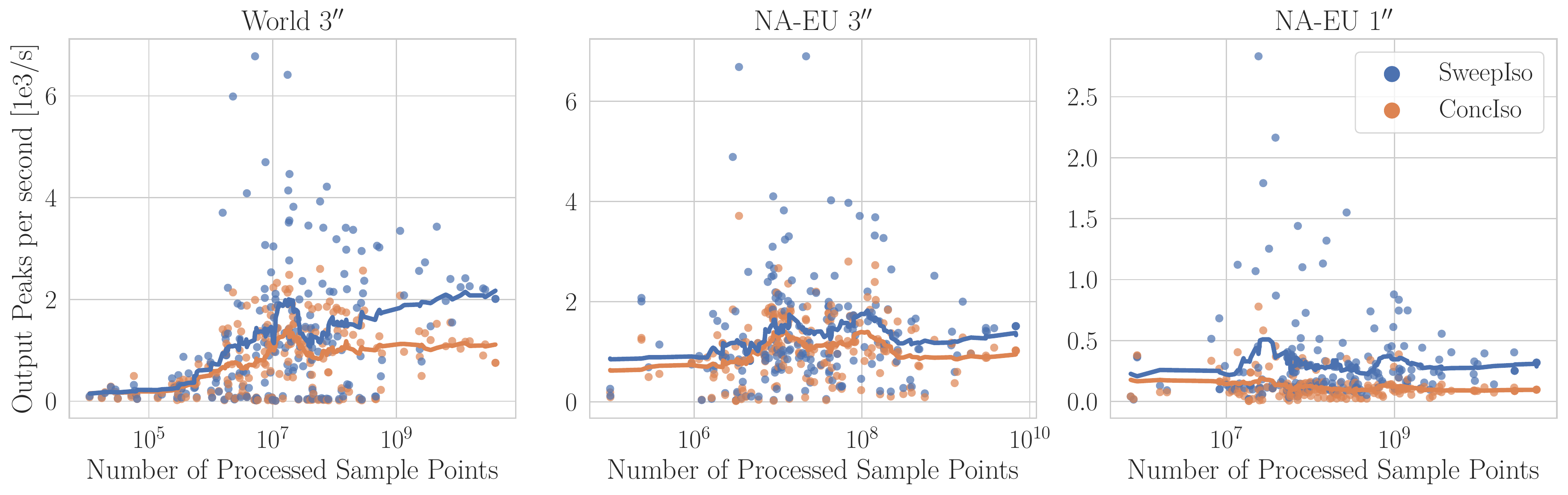}
		\caption{Throughput in output peaks per second over processed sample points.}
	\end{subfigure}
\end{figure}

\begin{figure}[tbh]\ContinuedFloat
	\centering
	\begin{subfigure}{\textwidth}
		\centering
		\includegraphics[width=\textwidth]{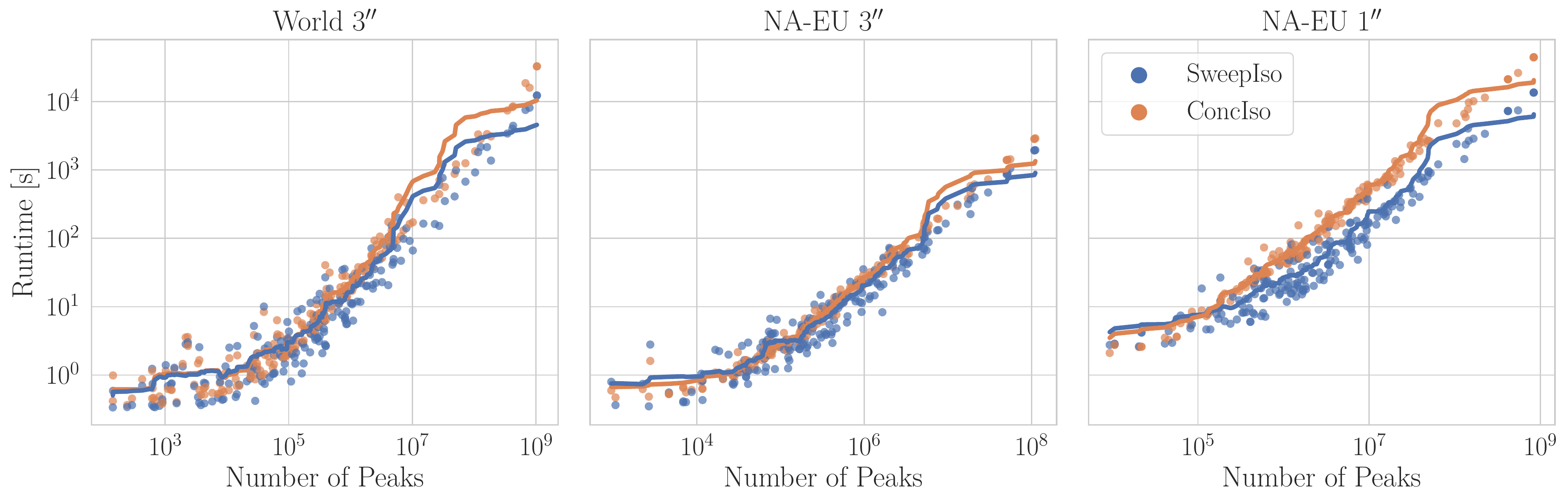}
		\caption{Execution time over processed peaks.}
	\end{subfigure}
	\begin{subfigure}{\textwidth}
		\centering
		\includegraphics[width=\textwidth]{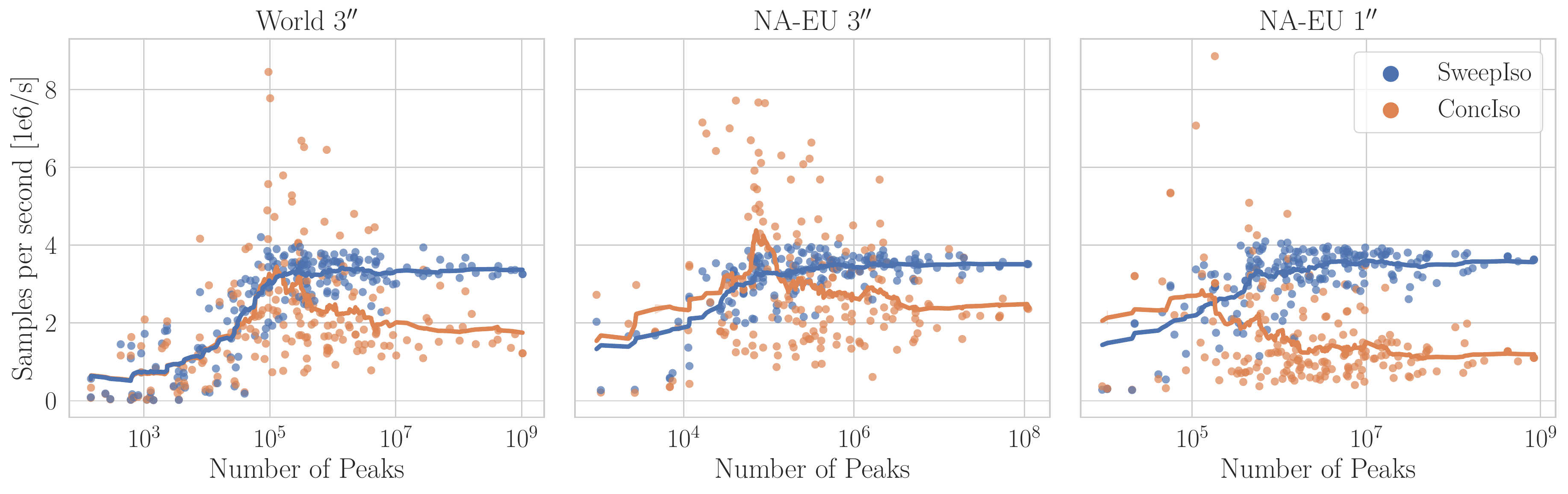}
		\caption{Throughput in points per second over processed peaks.}
	\end{subfigure}
	\begin{subfigure}{\textwidth}
		\centering
		\includegraphics[width=\textwidth]{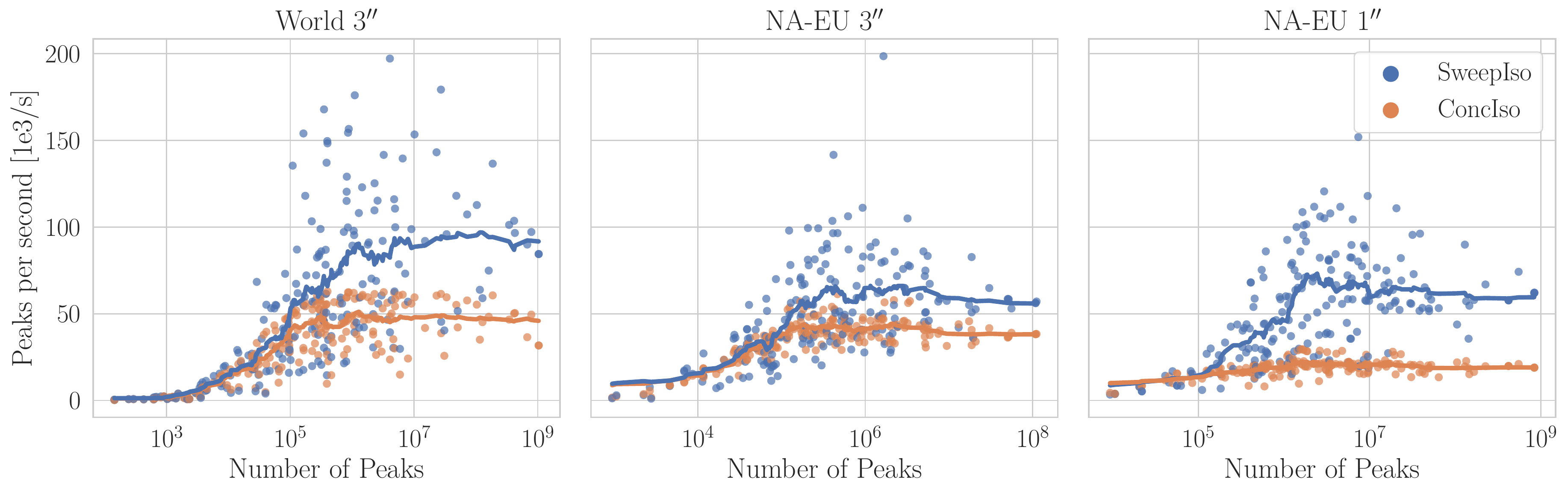}
		\caption{Throughput in peaks per second over processed peaks.}
	\end{subfigure}
	\begin{subfigure}{\textwidth}
		\centering
		\includegraphics[width=\textwidth]{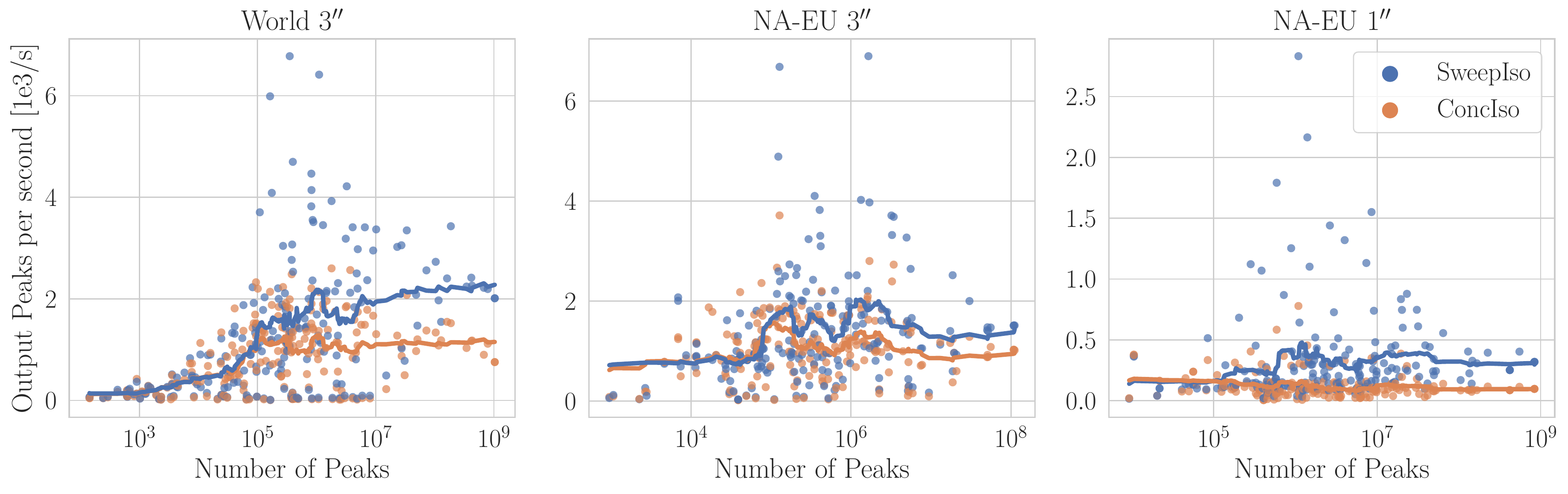}
		\caption{Throughput in output peaks per second over processed peaks.}
	\end{subfigure}
\end{figure}

\begin{figure}[tbh]\ContinuedFloat
	\centering
	\begin{subfigure}{\textwidth}
		\centering
		\includegraphics[width=\textwidth]{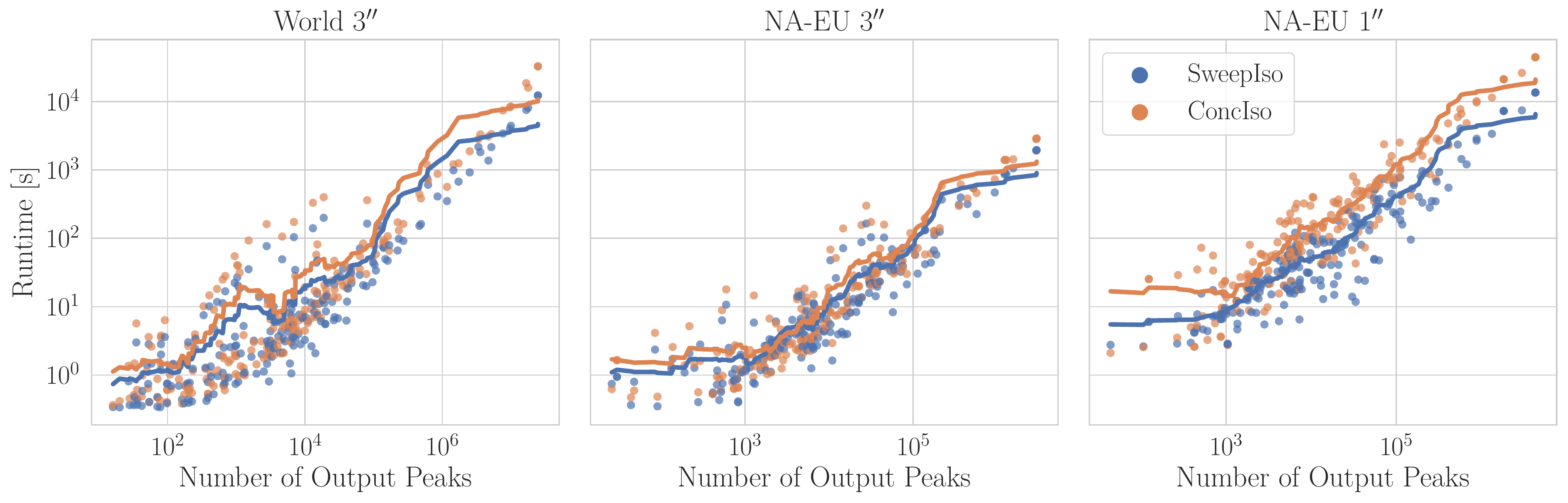}
		\caption{Execution time over output peaks.}
	\end{subfigure}
	\begin{subfigure}{\textwidth}
		\centering
		\includegraphics[width=\textwidth]{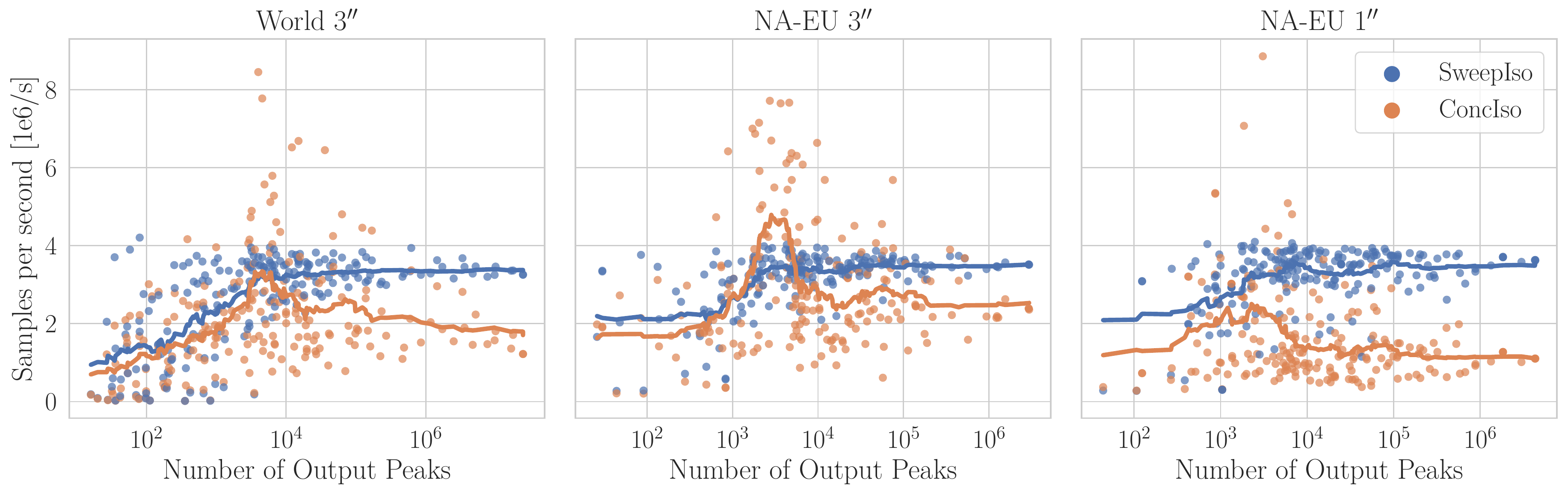}
		\caption{Throughput in points per second over output peaks.}
	\end{subfigure}
	\begin{subfigure}{\textwidth}
		\centering
		\includegraphics[width=\textwidth]{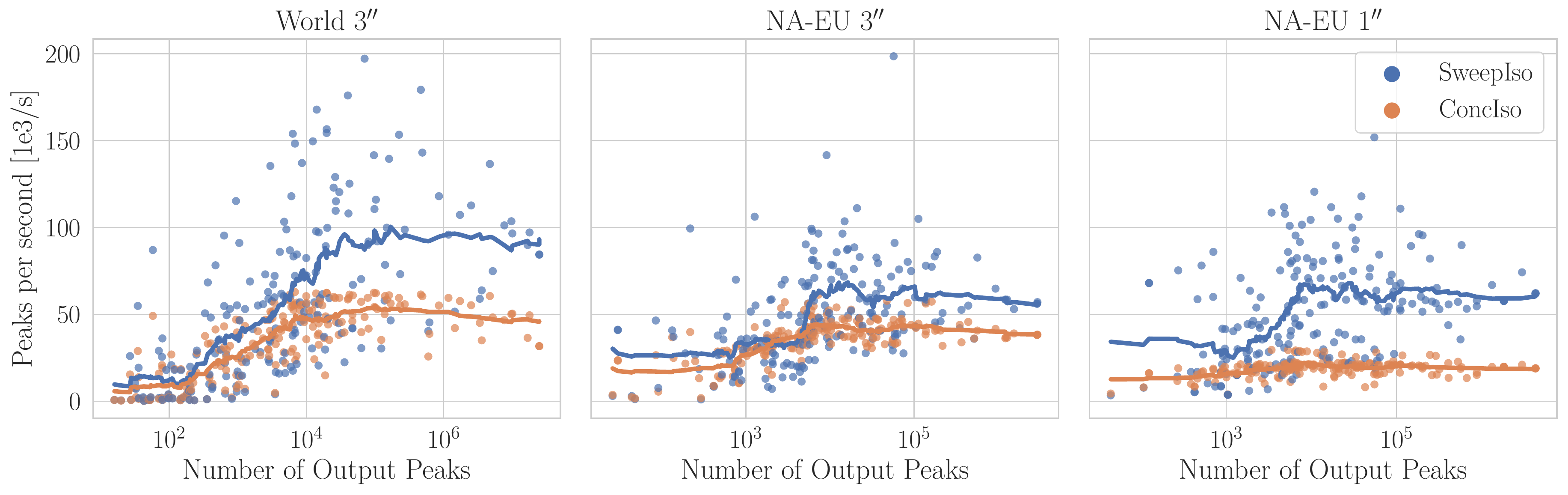}
		\caption{Throughput in peaks per second over output peaks.}
	\end{subfigure}
	\begin{subfigure}{\textwidth}
		\centering
		\includegraphics[width=\textwidth]{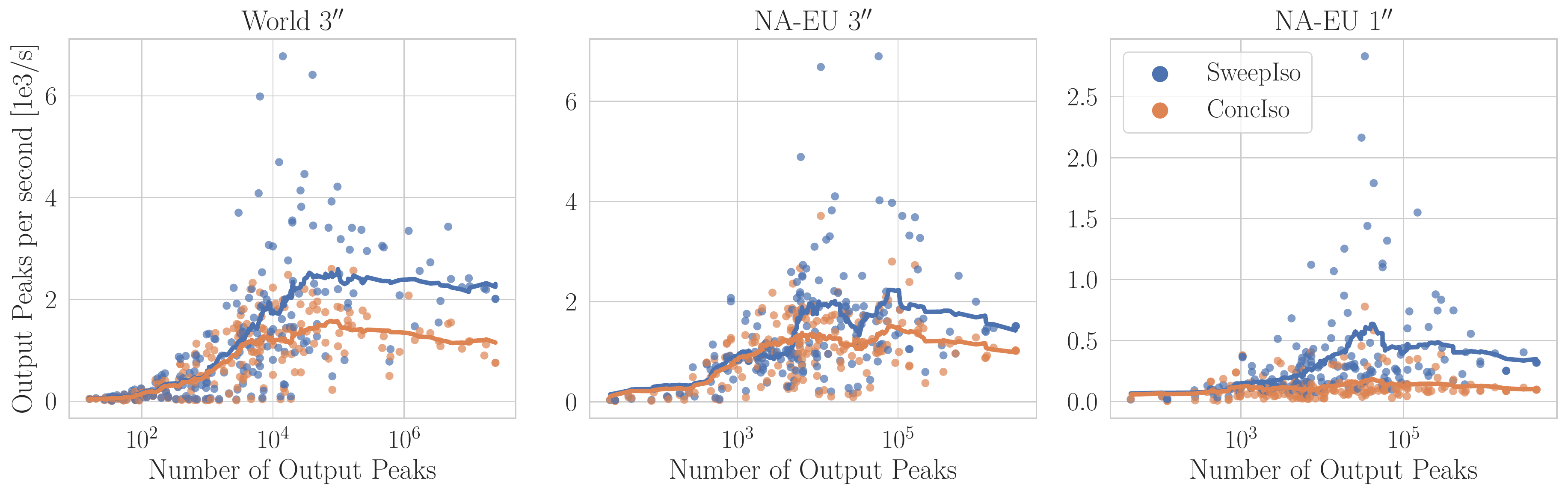}
		\caption{Throughput in output peaks per second over output peaks.}
	\end{subfigure}
\end{figure}

\begin{figure}[tbh]\ContinuedFloat
	\centering
	\begin{subfigure}{\textwidth}
		\centering
		\includegraphics[width=\textwidth]{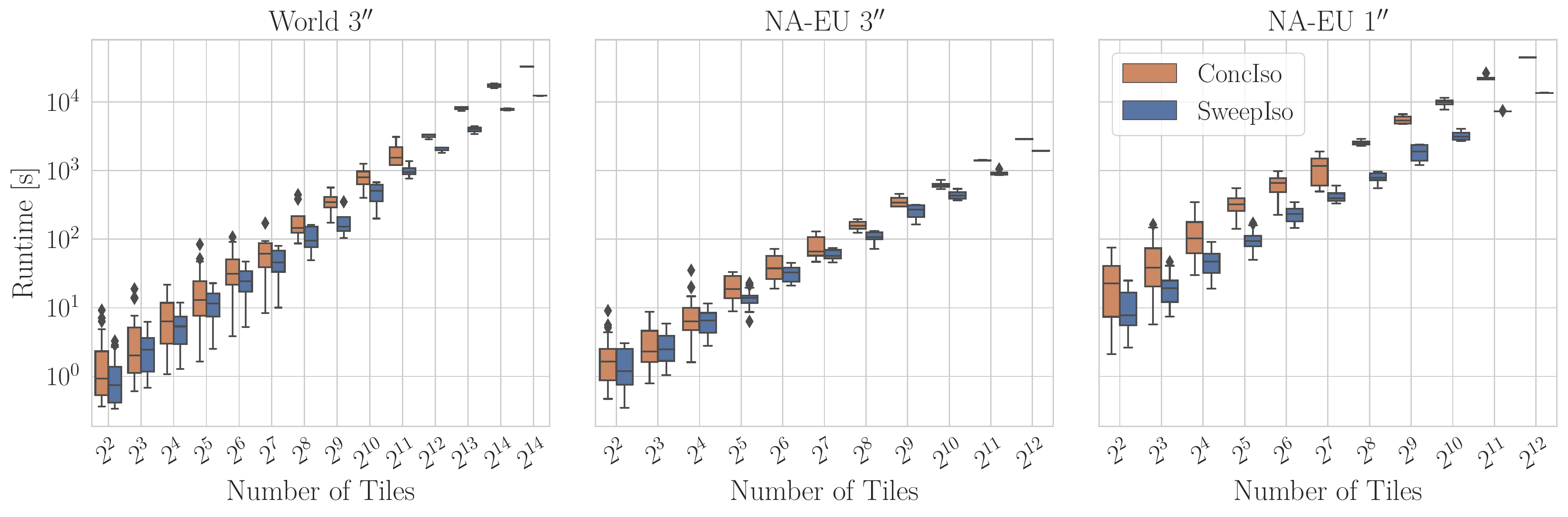}
		\caption{Execution time over number of tiles.}
	\end{subfigure}
	\begin{subfigure}{\textwidth}
		\centering
		\includegraphics[width=\textwidth]{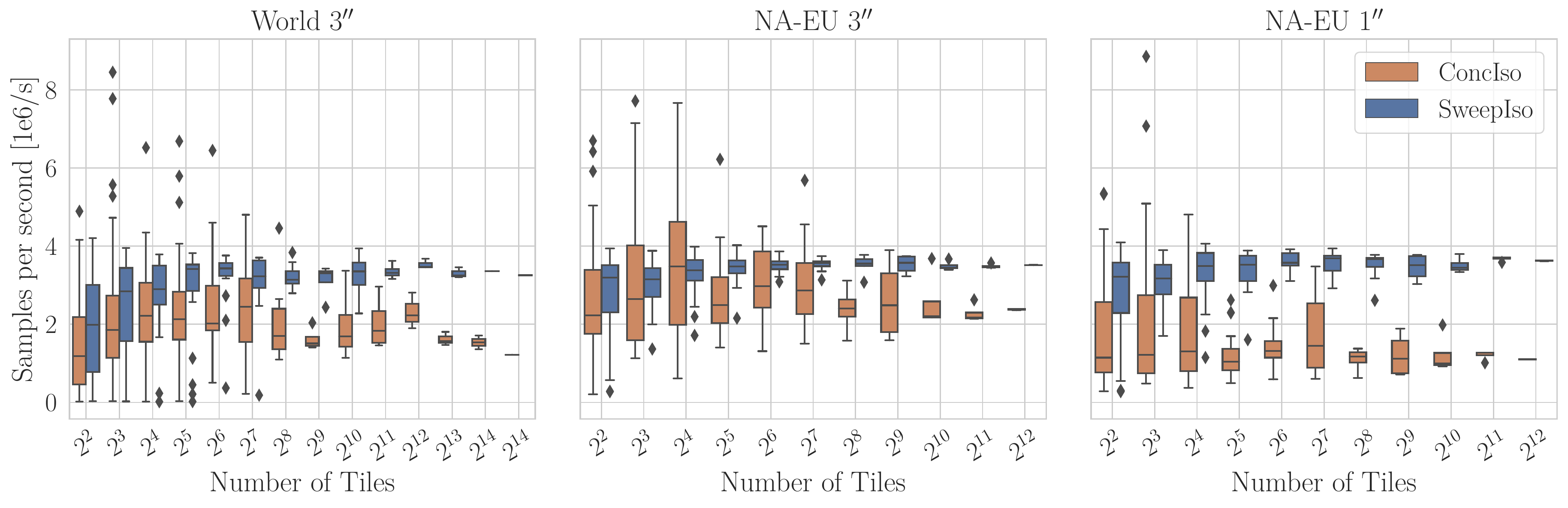}
		\caption{Throughput in points per second over number of tiles.}
		\label{fig:app:runtime:pointps}
	\end{subfigure}
	\begin{subfigure}{\textwidth}
		\centering
		\includegraphics[width=\textwidth]{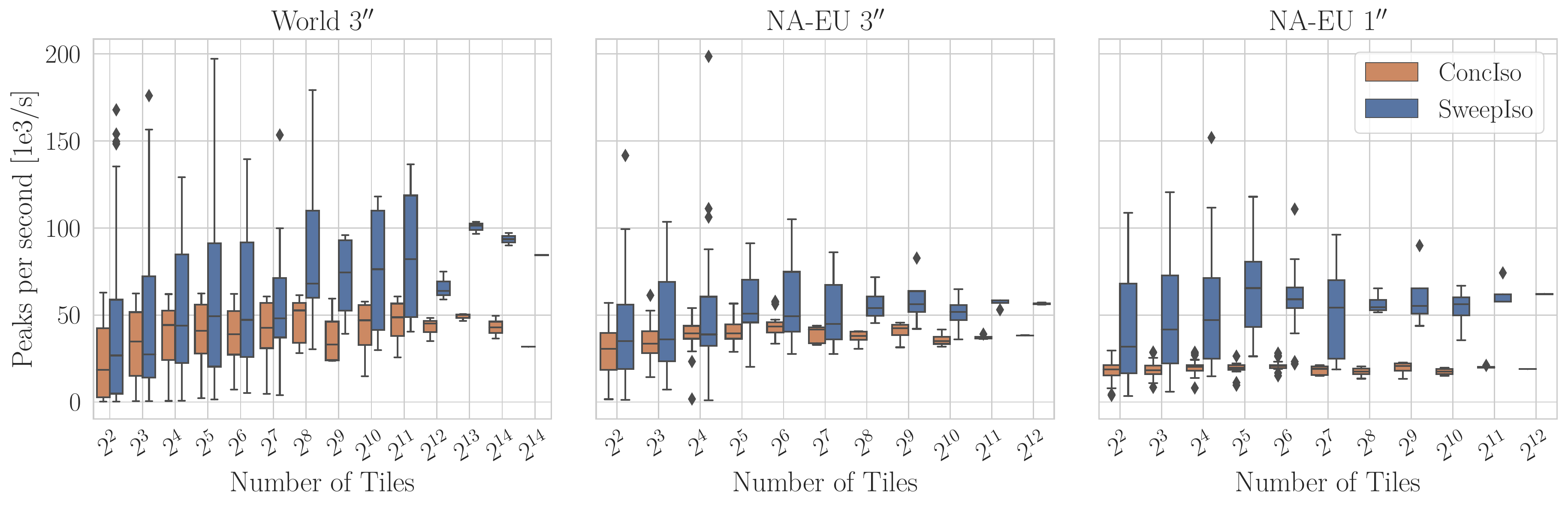}
		\caption{Throughput in peaks per second over number of tiles.}
		\label{fig:app:runtime:peaksps}
	\end{subfigure}
	\begin{subfigure}{\textwidth}
		\centering
		\includegraphics[width=\textwidth]{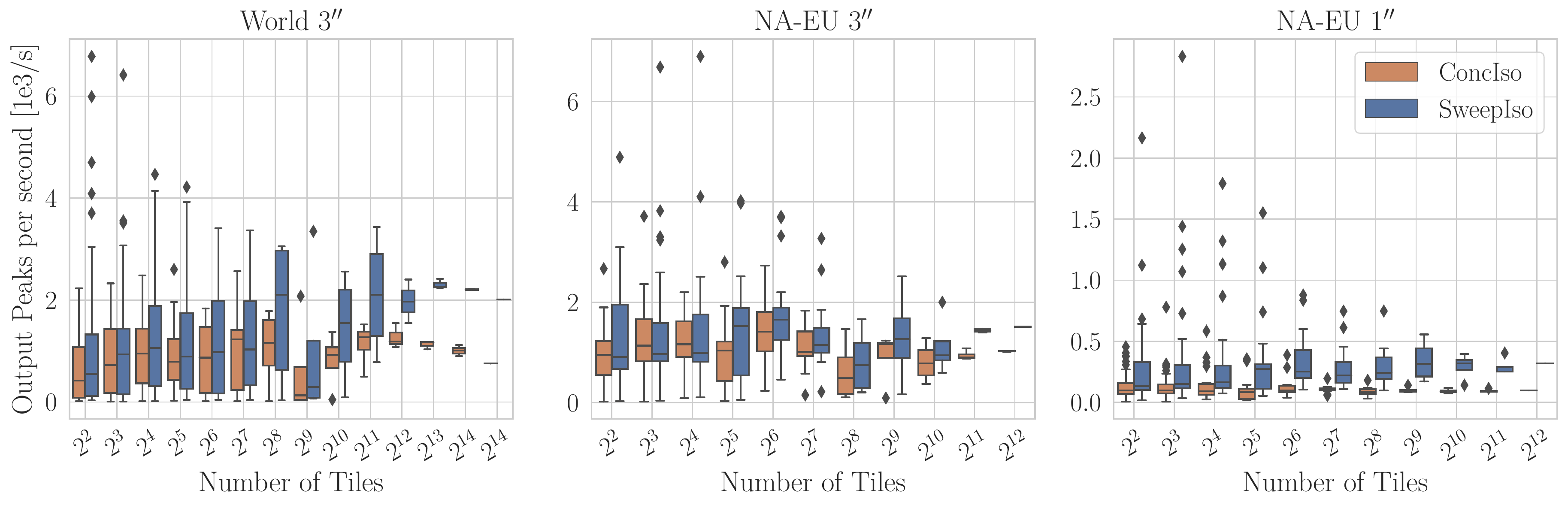}
		\caption{Throughput in output peaks per second over number of tiles.}
	\end{subfigure}
	\caption{Single-threaded runtime comparison of \ourAlg and \kirmseAlg.}
	\label{fig:app:runtime}
\end{figure}

\begin{figure}[tbh]
	\centering
	\includegraphics[width=\textwidth]{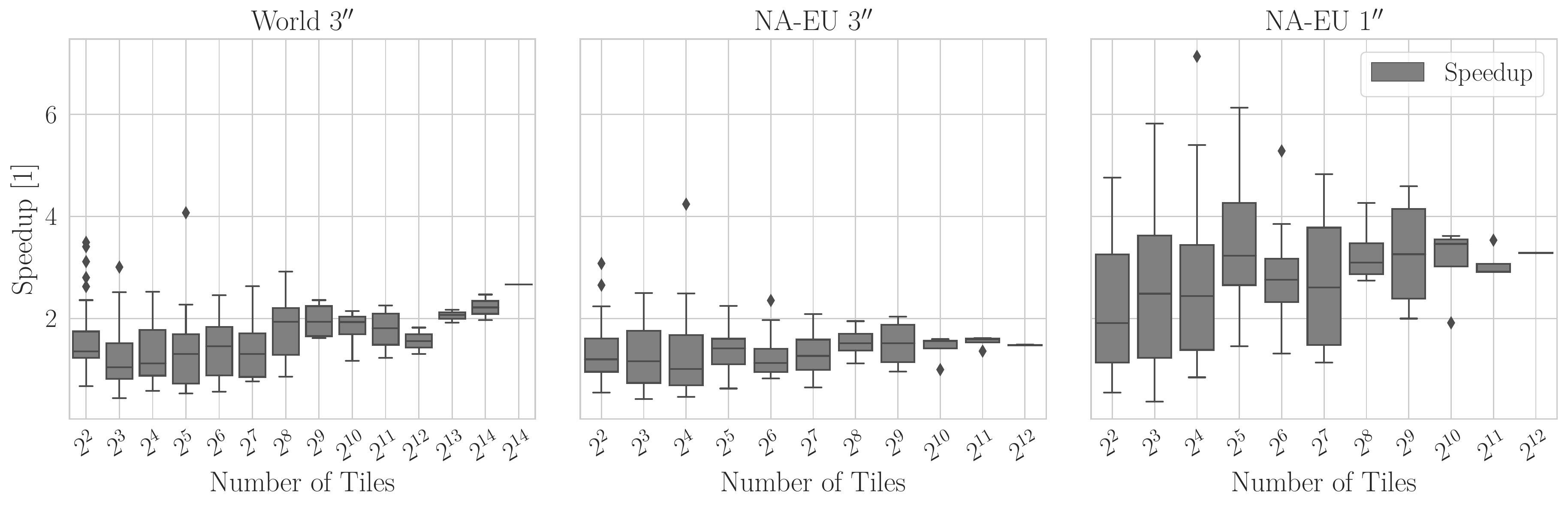}
	\caption{Single-threaded speedup comparison of \ourAlg and \kirmseAlg.}
	\label{fig:app:speedup}
\end{figure}

\end{document}